\documentclass[aps,prl,reprint,groupedaddress,showpacs,amsmath,amssymb,twocolumn,floatfix]{revtex4-1}
\usepackage{graphicx}
\usepackage{bm}
\usepackage{color}
\usepackage[normalem]{ulem}
\usepackage{cleveref}
\usepackage{amsmath}

\newcommand{\bk}{{\bf{k}}}
\newcommand{\bq}{{\bf{q}}}
\newcommand{\bp}{{\bf{p}}}
\newcommand{\br}{{\bf{r}}}

\newcommand{\eps}{\epsilon}

\begin{document}

\title{Tuning electron correlation in magic-angle twisted bilayer graphene using Coulomb screening}

\author{Xiaoxue Liu$^{1}$}
\author{Zhi Wang$^{1}$}
\author{K. Watanabe$^{2}$}
\author{T. Taniguchi$^{2}$}
\author{Oskar Vafek$^{3,4}$}
\author{J.I.A. Li$^{1}$$^{\dag}$}
\email{jia\_li@brown.edu}

\affiliation{$^{1}$Department of Physics, Brown University, Providence, RI 02912, USA}
\affiliation{$^{2}$National Institute for Materials Science, 1-1 Namiki, Tsukuba 305-0044, Japan}
\affiliation{$^{3}$Department of Physics, Florida State University, Tallahassee, FL 32306, USA}
\affiliation{$^{4}$ National High Magnetic Field Laboratory, Tallahassee, Florida, 32310, USA}

\date{\today}

\maketitle

\textbf{The ability to control the strength of interaction is essential for studying quantum phenomena emerging from a system of correlated fermions. For example, the isotope effect illustrates the effect of electron-phonon coupling on superconductivity, providing an important experimental support for the BCS theory  ~\cite{Maxwell1950,Bardeen1957}. %In addition, the ability to tune pairing strength in a fermionic cold atom system gives rise to a unique control of the crossover between the BEC and BCS regimes, uniting the strong and weak-pairing limits ~\cite{Randeria2014}. 
In this work, we report a new device geometry where the magic-angle twisted bilayer graphene (tBLG) is placed in close proximity to a Bernal bilayer graphene (BLG) separated by a $3$ nm thick barrier. Using charge screening from the Bernal bilayer, the strength of electron-electron Coulomb interaction within the twisted bilayer can be continuously tuned. Transport measurements show that tuning Coulomb screening has opposite effect on the insulating and superconducting states:  as Coulomb interaction is weakened by screening, the insulating states become less robust, whereas the stability of superconductivity is enhanced. Our results demonstrate the ability to directly probe the role of Coulomb interaction in magic-angle twisted bilayer graphene. Most importantly, the effect of Coulomb screening points toward electron-phonon coupling as the dominant mechanism for Cooper pair formation, and therefore superconductivity, in magic-angle twisted bilayer graphene. }

The discovery of superconductivity in magic-angle twisted bilayer graphene (tBLG) has raised intriguing questions regarding the nature of superconducting order parameter ~\cite{Cao2018b,Yankowitz2019SC,Lu2019SC}. The coexistence of the correlated insulator (CI) and superconducting phase is compared with cuprate materials, leading to suggestions that the superconducting phase arises from an unconventional origin ~\cite{Keimer2015hightc,Lee2006doping,Cao2018a,Cao2018b,Cao2020strange,Xu2018tblg,Isobe2018tblg,Guo2018tblg,Ray2019tblg,chichinadze2019nematic}. On the other hand, the more recent observations of superconductivity in the absence of correlated insulator appears to indicate that superconductivity arises through the electron-phonon coupling ~\cite{Stepanov2019interplay,Saito2019decoupling,Arora2020SC}, an interpretation that is backed by a range of theoretical models ~\cite{Ochi2018phonon,Lian2019phonon,Wu2018phonon}.  However, there is little agreement upon the origin of the superconducting states despite intense effort, owing to the lack of direct experimental evidence. 

It has long been recognized that the role of Coulomb interaction is essential to determining the nature of superconductivity. For a conventional superconductor, electron-phonon coupling competes against Coulomb repulsion in stabilizing superconductivity at low temperature ~\cite{McMillan1968transition,Allen1975transition}. As such, a weaker Coulomb repulsion will lead to more robust superconducting order parameters. On the other hand, an unconventional superconducting phase arises from an all-electron mechanism, where the order parameter strengthens with increasing Coulomb interaction ~\cite{Keimer2015hightc,Lee2006doping}. For conventional solid state materials, it remains an experimental challenge to directly control Coulomb interaction within a superconductor without introducing additional changes to the material.  The flexibility of the van der Waals technique offers a unique opportunity to control Coulomb interaction in magic-angle twisted bilayer structure using proximity screening ~\cite{Pizarro2019internal,Stepanov2019interplay,Saito2019decoupling}, which promises to shed light on the nature of superconductivity.

\begin{figure*}
\includegraphics[width=0.74\linewidth]{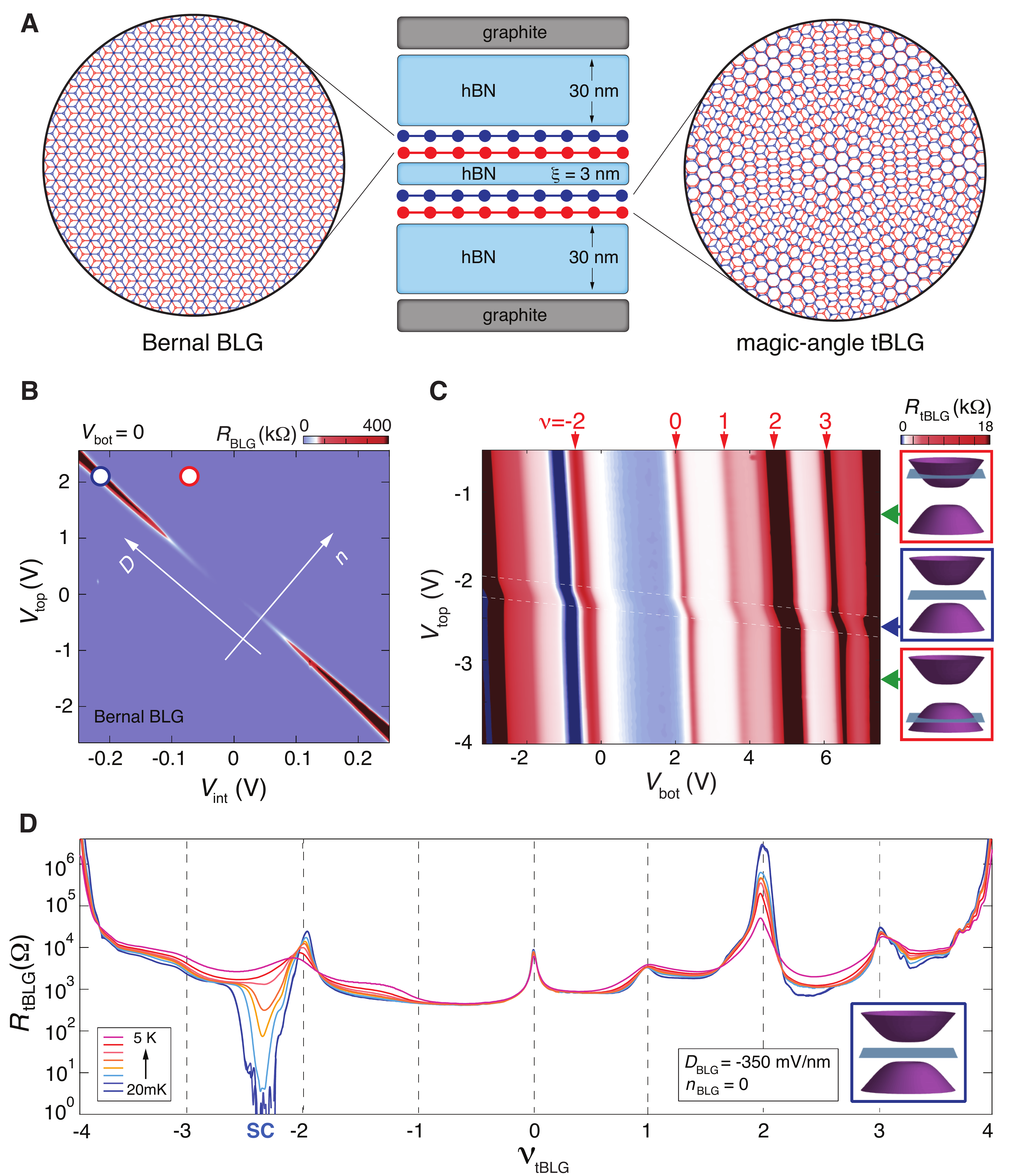}
\caption{\label{line} {\bf{Hybrid double-layer structure with Bernal BLG and tBLG}} (A) Schematic of the hybrid double-layer structure consisting of a Bernal BLG and a magic-angle tBLG, separated by a thin insulating barrier with thickness of $\xi = 3$ nm.  The structure is encapsulated with dual hexagonal boron nitride (hBN) dielectric and graphite gate electrodes.  (B) Longitudinal resistance of Bernal BLG $R_{BLG}$ as a function of $V_{top}$ and $V_{int}$. (C) Longitudinal resistance of tBLG $R_{tBLG}$ as a function of  $V_{top}$ and $V_{bot}$ measured at $T = 20$ mK and $D_{BLG} = -250$ mV/nm.  Screening from BLG is minimal between white dashed lines, where BLG is fulling insulating and tBLG is tuned with both top and bottom graphite gate, giving rise to the distortion in transport features. Inset: schematic energy structure of BLG at large $D_{BLG}$ with different $n_{BLG}$. (D) $R_{tBLG}$ as a function of filling fraction in tBLG $\nu$, measured at  $D = -350$ mV/nm and $n_{BLG}=0$ with varying temperature. 
%(E) Temperature dependence of $R_{tBLG}$ for the superconducting phase at optimal doping of $n_{tBLG} = -1.50 \times 10^{12}$ cm$^{-2}$. The dashed line corresponds to a fit to the normal state resistance, and the dotted line $50 \%$ of normal state resistance. Transition temperature is defined at the point where $R_{xx}$ is $50 \%$ the normal state resistance value, marked by the black arrow.  
}
\end{figure*}

A major road block for addressing the mechanism underlying the superconducting phase is the vastly different behavior between tBLG samples near the magic-angle, which may result from the spatial inhomogeneity of the moir\'e pattern ~\cite{Yoo2019atomic,Mcgilly2019seeing,Uri2019mapping}. The variation between different samples makes it difficult to provide reliable experimental constraints for a theoretical model. This obstacle is addressed in this work by tuning the strength of Coulomb interaction in a single device using screening, while studying the response in both the CIs and the superconducting phase using transport measurement.
We utilize a novel hybrid double-layer structure, where a Bernal bilayer graphene (BLG) and a magic-angle twisted bilayer graphene (tBLG) are separated by a thin insulating barrier with thickness of $\xi = 3$ nm, as shown in Fig.~1A. The close proximity allows charge carriers from BLG to screen Coulomb interaction in the tBLG, offering direct control on the nature of electron correlation in the moir\'e flat band. 

\begin{figure*}
\includegraphics[width=0.85\linewidth]{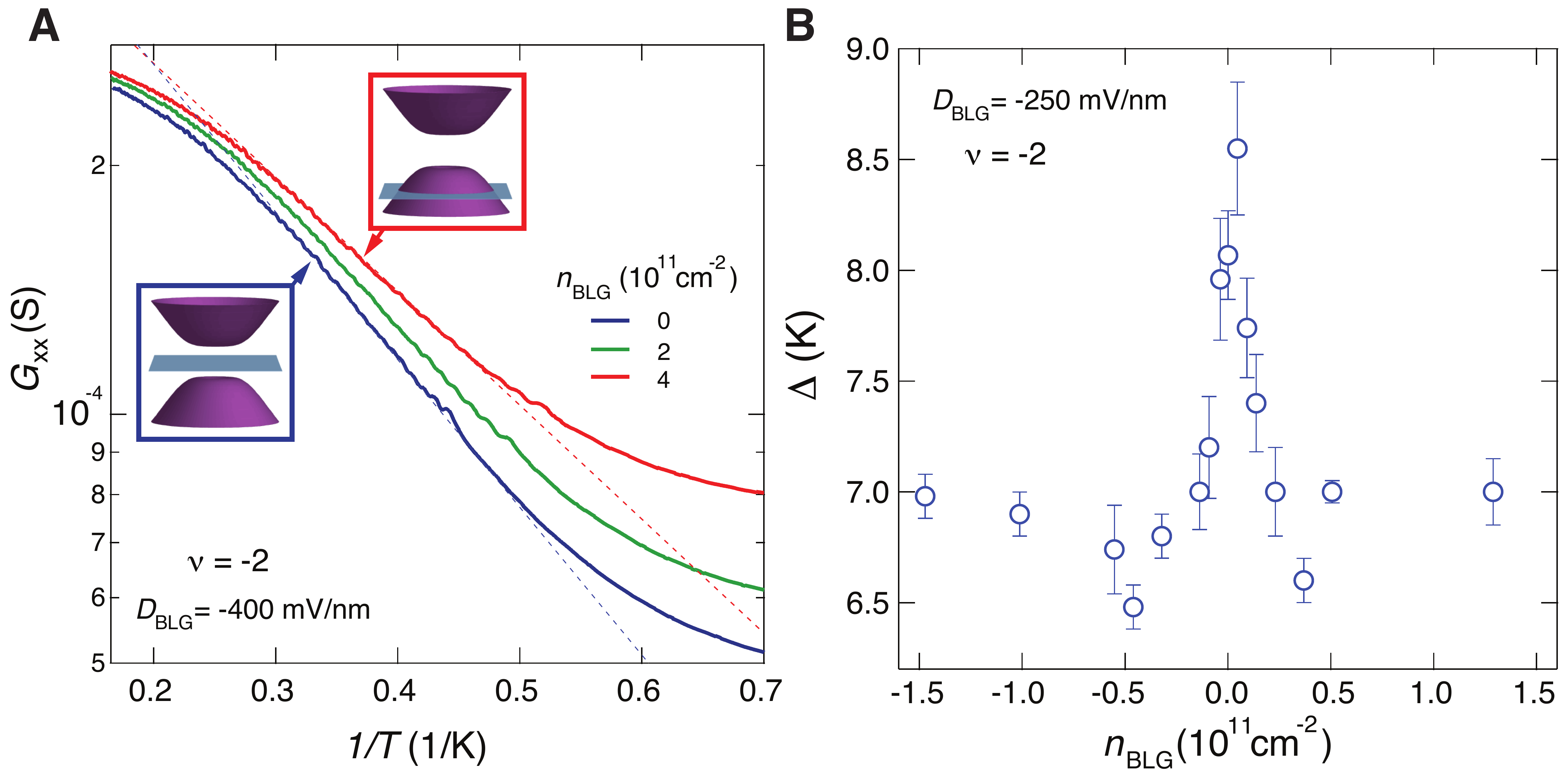}
\caption{\label{pattern}{\bf{The effect of Coulomb screening on the CI}}  (A) Arrhenius plot for the correlated insulating state at $\nu = -2$, measured at  $D_{BLG} = -400$ mV/nm with varying carrier density $n_{BLG}$. (B) Activation energy gap of the $\nu=-2$ CI state as a function of $n_{BLG}$. }
\end{figure*}

Charge carrier density in BLG and tBLG, $n_{BLG}$ and $n_{tBLG}$, can be independently controlled by varying the applied voltage to the top and bottom graphite gates, $V_{top}$ and $V_{bot}$ (see Method section for more details). Additionally, a voltage bias across two layers $V_{int}$ induces a perpendicular electric field, $D$, providing experimental control for the energy gap at the charge neutrality point in BLG ~\cite{Yuanbo.09,Li.17b,Zibrov2017}. Longitudinal resistance measured from the Bernal BLG displays a well-defined peak at the neutrality point which grows more insulating with increasing $D$-field, as shown in Fig.~1B. At large $D$-field, Bernal BLG can be tuned from fully insulating at $n_{BLG}=0$ (blue circle in Fig.~1B) to highly conductive at large $n_{BLG}$ (red circle in Fig.~1B), offering a large contrast in the strength of Coulomb screening.  Fig.~1C plots the transport response of tBLG, as $n_{BLG}$ and $n_{tBLG}$ are both tuned by varying $V_{top}$ and $V_{bot}$. The boundaries of the moir\'e flat band are defined by the emergence of insulating states at $n_{tBLG} = \pm 2.59 \times 10^{12}$ cm $^{-2}$, which corresponds to a twist angle of $\theta = 1.06^{o}$. A normalized density scale, marked by red arrows, is defined to describe partial filling $\nu$ of the moir\'e band based on the $4$-fold degeneracy of spin and valley degrees of freedom for both electron and hole type carriers.  

The static $e^2/r$ Coulomb interaction among the electrons with charge $e$, separated by distance $r$ in the tBLG is modified by the presence of BLG leading to their effective interaction energy 
\begin{eqnarray}\label{Eqn:Veff}
V^{eff}(\br)&=&\int \frac{d^2\bq}{(2\pi)^2} \frac{2\pi e^2}{|\bq|}\left[1-e^{-2|\bq|\xi}\left(1-\frac{1}{\eps_{AB}(\bq)}\right)\right]e^{i\bq\cdot\br}.\nonumber\\
\end{eqnarray}
The wavevector $\bq$ dependent dielectric constant of the BLG can be related to its static polarization function $\Pi^0_{\bq}$ as $\eps_{AB}({\bq})=1+\frac{2\pi e^2}{|\bq|}\Pi^0_{\bq}$. When the BLG is gated to a finite carrier density and thus acts as a metal, $\Pi^0_{{\bq}\rightarrow 0}\rightarrow const.$ and the $\eps_{AB}({\bq\rightarrow 0})$ diverges. At long distances $V^{eff}(\br)$ then corresponds to the real space potential produced by the test charge and its mirror image a distance $2\xi$ above the twisted bilayer. When the BLG is insulating, $\Pi^0_{{\bq}\rightarrow 0}\sim {\bq}^2$ and the $V^{eff}(\br)$ is unchanged at long distances by the presence of the BLG bilayer (see supplementary information for details of Coulomb screening in a hybrid double-layer structure at any $\br$).
Since the strength of Coulomb screening is correlated with the conductivity of BLG,
its effect can be studied by comparing transport properties of tBLG in- and outside the density range marked by white dashed lines in Fig.~1C. %As such, the hybrid double-layer structure provides a unique and versatile experimental platform to study and manipulate electron correlation in the magic-angle tBLG, which promises to shed light on the underlying mechanism of the superconducting phase. 

\begin{figure*}
\includegraphics[width=1\linewidth]{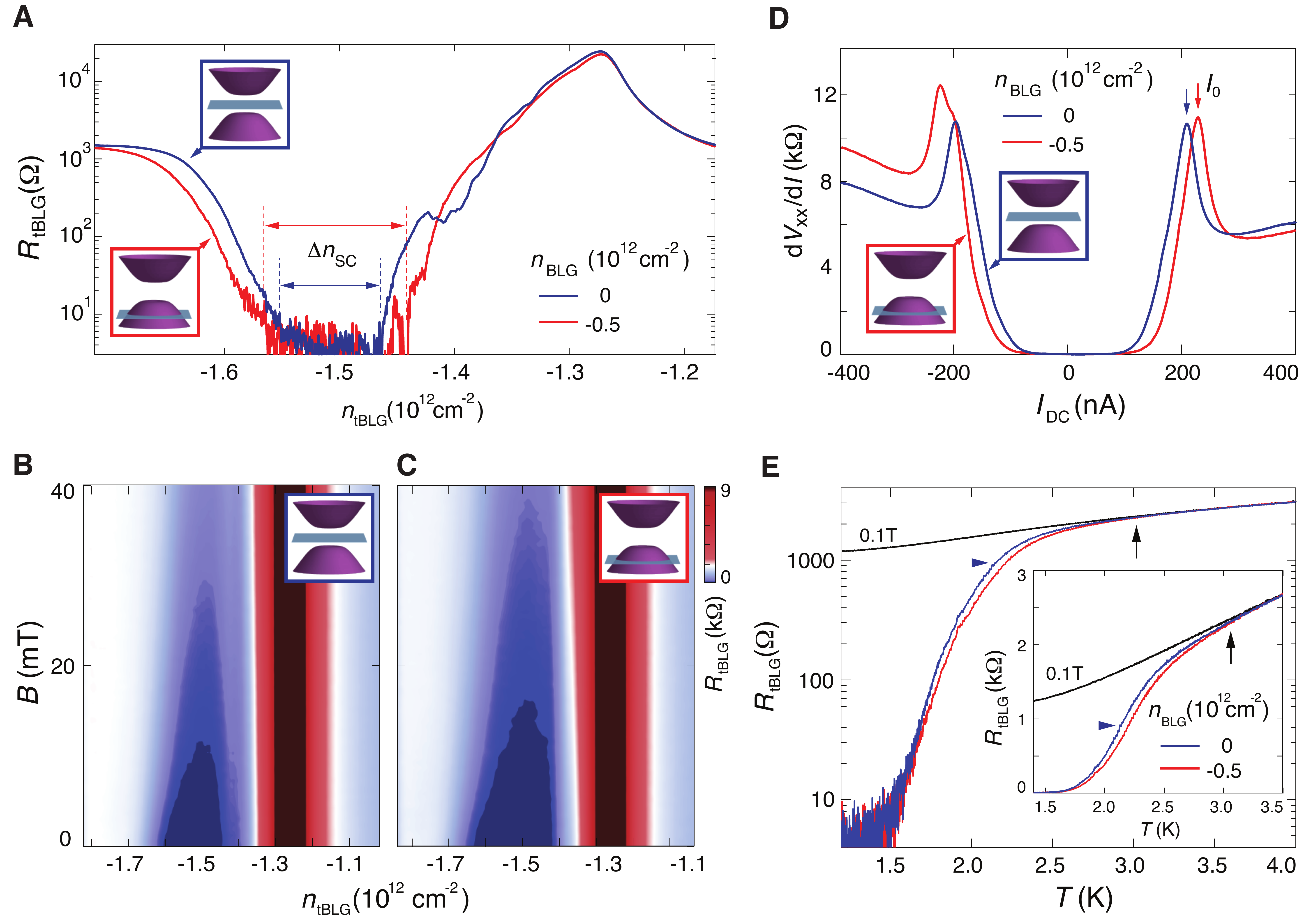}
\caption{\label{datamap} {\bf{Tuning $T_c$ of the superconducting phase}} The effect of tuning $n_{BLG}$ in the presence of a large $D$-field induced energy gap in BLG. (A) $R_{tBLG}$ as a function of carrier density in tBLG, $n_{tBLG}$, measured at $D = -350$ mV/nm with different $n_{BLG}$. (B-C)  $R_{tBLG}$ as a function of carrier density $n_{tBLG}$ and magnetic field $B$.  BLG is tuned to $D_{BLG} = -350$ mV/nm with (B) $n_{BLG}=0$ and (C) $n_{BLG}=-0.5 \times 10^{12}$ cm$^{-2}$.  (D) Differential resistance $dV_{xx}/dI$ versus d.c. bias current $I$ measured at  optimal doping  $n_{tBLG} = -1.5 \times 10^{12}$ cm$^{-2}$ and base temperature of $T = 20$ mK, with BLG tuned to different density. A critical current $I_o$ is defined by the peak in $dV_{xx}/dI$, where superconductivity is destroyed and the differential resistance goes over to the normal state value ~\cite{Benyamini2019fragility}.  (E) $R_{tBLG}$ as a function of temperature measured at optimal doping $n_{tBLG} = -1.50 \times 10^{12}$ cm$^{-2}$ and $D_{BLG} = -350$ mV/nm for different $n_{BLG}$. The blue and red traces are measured at $B=0$, while the black trace is measured at $B=0.1$ T where superconductivity is fully suppressed.  Inset shows the temperature dependence of $R_{tBLG}$ on a linear scale. $T_c$ is operationally defined by $50 \%$ of normal state resistance and marked by the blue horizontal arrow. The separation of $B=0$ and $B=0.1$ T curves marks the onset of pairing (vertical black arrow).}
\end{figure*} 

First, we examine the transport response of tBLG in the absence of Coulomb screening, by measuring inside the white dashed lines in Fig.~1C where BLG is fully insulating. 
Fig.~1D plots longitudinal resistance of tBLG, $R_{tBLG}$, over the full filling range of the moir\'e flat band. Apart from the charge neutral point (CNP), a series of resistive features emerge at $\nu = \pm 2$, $+1$ and $+3$, which are consistent with the correlated insulating (CI) states from previous observation ~\cite{Cao2018b,Yankowitz2019SC,Lu2019SC}. In addition,
a robust regime of superconducting phase emerges at low temperature, evidenced by $R_{xx}$ dropping to zero. 
The robust CI and superconducting states establish an excellent starting point, allowing us to quantitatively examine the effect of Coulomb screening by studying the variation in transport behavior while varying $n_{BLG}$. 
%At the optimal doping of $n_{tBLG} = -1.50 \times 10^{12}$ cm$^{-2}$,
%Fig.~2d plots the temperature dependence of the superconducting phase, showing 
%the transition temperature $T_c$, operationally defined as $50 \%$ of extrapolated normal state resistance, is shown to be  $2.17$ K in Fig.~1E, which is in line with previous observation in tBLG with similar twist angle ~\cite{Saito2019decoupling}.

\begin{figure*}
\includegraphics[width=1\linewidth]{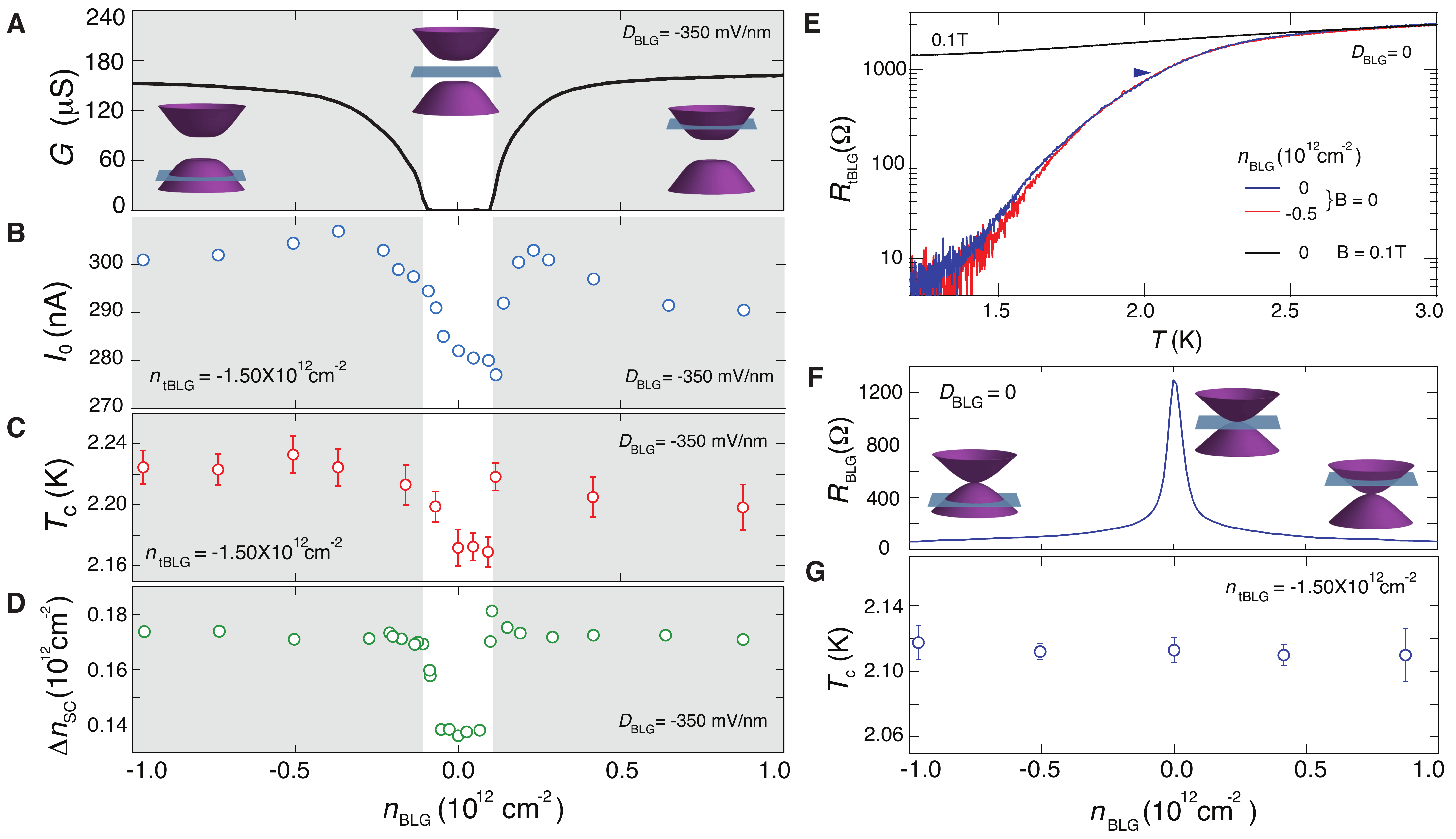}
\caption{\label{fig4} {\bf{$B$-field and current dependence, normal state resistance }}    (A) $G_{xx}$ in BLG as a function of carrier density $n_{BLG}$ measured at $D_{BLG} = -350$ mV/nm. Insets show the position of fermi level relative to the energy band structure in BLG at different carrier density.
(B) $I_o$, (C) 
$T_c$ and (D) $\Delta n_{SC}$ as a function of carrier density in BLG, $n_{BLG}$, measured at $D_{BLG} = -350$ mV/nm. $I_o$ and $T_c$ are measured at optimal doping $n_{tBLG} = -1.50 \times 10^{12}$ cm$^{-2}$.  A large energy gap is induced in BLG by $D$-field. (E) $R_{tBLG}$ as a function of temperature measured at optimal doping $n_{tBLG} = -1.50 \times 10^{12}$ cm$^{-2}$ and $D_{BLG} = 0$ for different $n_{BLG}$. The blue and red traces are measured at $B=0$, while the black trace is measured at $B=0.1$ T where superconductivity is fully suppressed. Tuning $n_{BLG}$ gives rise to small variations in $R_{tBLG}$ at temperature much lower than $T_c$, while no effect is observed in temperature range near $T_c$. (F) $R_{xx}$ in BLG as a function of carrier density $n_{BLG}$ measured at $D = 0$. (G) $T_c$ as a function of carrier density in BLG, $n_{BLG}$, measured at $D_{BLG} = 0$ and optimal doping $n_{tBLG} = -1.50 \times 10^{12}$ cm$^{-2}$.
}
\end{figure*}

Fig.~2A plots the temperature dependence of 4-terminal conductance of the CIs at $\nu=-2$, which exhibits thermally activated behavior with strong $n_{BLG}$ dependence.  An energy gap $\Delta$ can be extracted from the slope of the Arrhenius plot (dashed lines in Fig.~2A), providing a measure for the strength of the CI state. The effect of Coulomb screening is investigated by measuring the energy gap $\Delta$ while varying density in BLG $n_{BLG}$, as shown in Fig.~2B. $\Delta_{\nu=-2}$ is the largest when BLG is fully insulating at $n_{BLG}=0$. 
Similar behavior in energy gap is observed at $\nu=2$ and $3$ as a function of $n_{BLG}$ (see SI for energy gap measurements of CIs at $\nu=2$ and $3$).  Since CIs at integer filling arise from Coulomb correlation within the moir\'e flat band ~\cite{Bistritzer2011,Wu2007flat,Cao2018a,KangVafekPRL2019}, the trend in  $\Delta$ provides strong evidence that electron correlation in tBLG is directly tunable using Coulomb screening: screening from the BLG decreases as BLG becomes insulating, leading to stronger Coulomb interaction and a larger $\Delta$ for the CI states. In addition,  a minimum in $\Delta_{\nu=-2}$ is observed near the band edge of BLG at $n_{BLG} \sim 5 \times 10^{11}$ cm$^{-2}$, suggesting that Coulomb screening is the strongest when the fermi level of BLG is near the van Hove singularities where the density of states is the largest ~\cite{Young2011capacitance,Hunt.17}. We note that BLG at $n_{BLG} = 0$ remains insulating  over the temperature range of the thermal activation measurement, ensuring the strength of Coulomb screening remains a constant with varying temperature and is controlled only by $n_{BLG}$  (see SI for temperature dependence of transport response in BLG). The fact that the CI at $\nu=-2$ remains robust in the presence of strong Coulomb screening suggests that CIs cannot be fully suppressed by Coulomb screening alone ~\cite{Stepanov2019interplay}. 
The effect of Coulomb screening on CIs shows excellent agreement with theoretical models (see SI for more detailed theoretical discussion), which provides an important reference for studying the nature of superconductivity in tBLG using Coulomb screening from BLG.

Having established the effect of Coulomb screening on CIs, we turn our attention to the superconducting phase on the hole doping side of $\nu=-2$. Fig.~3A plots $R_{tBLG}$ at $20$mK as a function of carrier density in tBLG, $n_{tBLG}$, demonstrating an apparent effect when tuning carrier density in BLG, $n_{BLG}$: 
%demonstrating that superconductivity is stable over density range defined by zero resistance. 
the superconducting phase is stable over a wider density range when BLG is metallic (red trace in Fig.~3A), as compared to when BLG is fully insulating (blue trace in Fig.~3A). The boundary of the superconducting region is defined by the density where $R_{tBLG}$ increases above the noise level, which determines the density range of the superconducting dome, $\Delta n_{SC}$, as shown in Fig.~3A.
The effect of $n_{BLG}$-tuning is also observed in the density-magnetic field phase diagram, where zero resistance is shown as dark blue in the chosen color scale. A larger superconducting region is observed at large $n_{BLG}$ (Fig.~3C) compared to $n_{BLG}=0$ (Fig.~3B).  Fig.~3D and 3E show the measurements of the  critical current $I_o$ and critical temperature $T_c$ of the superconducting phase, which confirm the same trend: the superconducting phase is more robust when BLG is metallic compared to fully insulating. %At the optimal doping of $n_{tBLG} = -1.50 \times 10^{12}$ cm$^{-2}$,
%Fig.~2d plots the temperature dependence of the superconducting phase, showing 
We note that $R_{tBLG}$ measured at $B=0.1$ T, where superconductivity is fully suppressed, reflects the transport behavior of tBLG in the normal state. 
The transition temperature $T_c$, operationally defined as $50 \%$ of extrapolated normal state resistance, is shown to be  $\sim 2.2$ K in Fig.~3E, which is in line with previous observation in tBLG with similar twist angle ~\cite{Saito2019decoupling}.
%Fig.~3E plots $R_{tBLG}$ measured at $B=0.1$ T, where superconductivity is fully suppressed, 
The temperature dependence of $R_{tBLG}$ leads to a few important observations: (i) the normal state $(T\gtrsim 3K)$ resistance in tBLG is insensitive to changes in BLG, demonstrating that modification in the impurity scattering resulting from a nearby metallic layer does not play a dominating role ~\cite{Ponomarenko2011tunable}; (ii) the onset of Cooper pairing is observed at $T \sim 3$ K (vertical black arrow in Fig.~3E), evidenced by the bifurcation between $R_{tBLG}$ measured at $B=0$ and $0.1$T. Because the $n_{BLG}$-tuning onsets in a similar temperature range as the onset of Cooper pairing, as shown by the red and blue traces in Fig.~3E, we conclude that $n_{BLG}$-tuning has a direct impact on the mechanism of Cooper pair formation in tBLG. 

To further demonstrate the effect of $n_{BLG}$-tuning, we examine the stability of the superconducting phase by measuring $I_o$, $T_c$ and $\Delta n_{SC}$ as a function of $n_{BLG}$.
The values of all three parameters closely follow the conductance of BLG, as shown in Fig.~4, with the superconducting phase being more robust when BLG is metallic. 
This trend allows us to characterize the coupling between BLG and tBLG. In the presence of external radiation, the superconducting phase is significantly suppressed when BLG is metallic, possibly resulting from capacitive coupling between BLG and tBLG across the thin insulating barrier ~\cite{Kellogg2005}. Such coupling is suppressed when BLG is fully insulating, or with an external low-pass filtering (see SI for more details regarding low-pass filtering) ~\cite{Tamir2019sensitivity}.  The fact that superconductivity is enhanced by a nearby metallic layer indicates that the influence of an external radiation is sufficiently eliminated by the low-pass filter. 

The enhancement of superconductivity by a nearby metallic BLG layer could, in principle, result from suppressed phase fluctuations in tBLG, without a significant effect on Cooper pair formation, giving rise to a higher Berezinskii-Kosterlitz-Thouless transition temperature $T_{BKT}$  ~\cite{Merchant2001crossover}. We can rule out this scenario because the $n_{BLG}$-tuning onsets as does the Cooper pairing (Fig.~3E). 
Additional support for this conclusion is provided by our data at $D_{BLG}=0$, which closes the gap at the neutrality point of BLG and makes it conducting, albeit with resistivity varying by about a factor of $20$ between the CNP and away from it (Fig.~4F).  Although there is an effect on phase fluctuations i.e. once the resistance drops by an order of magnitude below its normal state value, there is no effect on the Cooper pairing scale as demonstrated in the Fig~4E, or on $T_c$ (Fig.~4G).  Notwithstanding the above-mentioned, we clearly observe that the value $T_{BKT}\approx 1.4$K is significantly below the pairing onset temperature $\sim 3$K, possibly pointing to a wide window of paring fluctuations.  Here $T_{BKT}$ is defined by fitting $R_{tBLG}$ using the Halperin-Nelson form (see Fig.~SI\ref{fig:HN}) ~\cite{Lin2012BKT}.

%While tuning $n_{BLG}$ gives rise to small variations in $R_{tBLG}$ at temperature much lower than $T_c$, which can be attributed to changes in phase fluctuations, there is no effect on the Cooper pairing scale as demonstrated in the Fig~4E, or on $T_c$ (Fig.~4G).

%(*DO WE NEED THIS?*)
%Due to the close proximity between Bernal BLG and tBLG, the influence of BLG on the stability of the superconducting phase in tBLG could result from a few possible mechanisms. 
%While multiple pockets are observed in some previous observations,  ~\cite{Cao2018b,Yankowitz2019SC,Lu2019SC,Stepanov2019interplay}. This discrepancy likely results from variation in sample details

%We note that only one superconducting pocket is observed in our sample on the hole doping side of $\nu=-2$, which is similar to previous observations in reference ~\cite{Saito2019decoupling} where Coulomb screening from nearby graphite layer is weak. Based on this, the absence of other superconducting pockets is likely associated with sample details, such as twist angle inhomogeneity, and we do not think it results from Coulomb screening from a nearby BLG.
%(*END of "DO WE NEED THIS?"*)

Lastly, it is worth pointing out that changes in Coulomb screening do not influence the linear-in-$T$ behavior in tBLG at high temperature (see Fig.~SI\ref{fig:slope}). It has been suggested that the slope of $R_{tBLG}$ in the $T$-linear regime is associated with the strength of quasielastic electron scattering off acoustic phonon modes ~\cite{Wu2019phonon,Polshyn2019phonon}. Transport measurement in this regime demonstrates a constant slope of $R_{tBLG}$ independent of carrier density in BLG, $n_{BLG}$ (see Fig.~SI\ref{fig:slope}), which would imply that the strength of acoustic electron-phonon coupling is not influenced by Coulomb screening. Taken together, our result suggests that the dominant effect of $n_{BLG}$-tuning is changing the strength of Coulomb interactions in the magic-angle tBLG. The fact that Cooper pair formation and superconductivity become more robust with increasing screening is consistent with electron-phonon coupling competing against Coulomb interaction to stabilize the superconducting phase ~\cite{McMillan1968transition}. The ability to control Coulomb interaction promises to provide important constraints for theoretical models aiming to accurately describe superconductivity in magic-angle tBLG.

\section*{Acknowledgments}
We thank M. Yankowitz, A. F. Young, C. R. Dean Q. Shi, S. Todadri, S. A. Kivelson for discussions and Y. Zeng for helpful input on fabrication. This work was primarily supported by Brown University. Device fabrication was performed in the Institute for Molecular and Nanoscale Innovation at Brown University. O.~V.~was supported by NSF DMR-1916958, and by the National High Magnetic Field Laboratory through NSF Grant No.~DMR-1157490 and the State of Florida. K.W. and T.T. acknowledge support from the EMEXT Element Strategy Initiative to Form Core Research Center, Grant Number JPMXP0112101001
and the CREST(JPMJCR15F3), JST.

\section*{Competing financial interests}
The authors declare no competing financial interests.

\newpage

\newpage

\begin{widetext}
\section{supplementary material}
\renewcommand{\figurename}{Fig.SI}
\setcounter{figure}{0}

\subsection{Screening of the Coulomb interaction due to the Bernal bilayer}

The interaction among the electrons in the twisted bilayer is modified due to the presence of the Bernal bilayer.
The Hamiltonian for the combined system is
\begin{eqnarray}
H_{int}&=& \int d^2{\br}\int d^2{\br'} \left( \frac{1}{2}n_1({\br})V^\parallel({\br}-{\br'})n_1({\br}')
+ \frac{1}{2}n_2({\br})V^\parallel({\br}-{\br'})n_2({\br}')+ n_1({\br})V^\perp({\br}-{\br}')n_2({\br}')\right),
\end{eqnarray}
where ${\br}$ and ${\br}'$ are 2D position vectors, $n_1(\br)$ is the electron density in the twisted bilayer, and $n_2(\br)$ is the electron density in the Bernal bilayer. The intra-layer Coulomb interaction is
\begin{eqnarray}
V^{\parallel}(\br)&=&\frac{e^2}{|\br|}=\int \frac{d^2\bq}{(2\pi)^2} \frac{2\pi e^2}{|\bq|}e^{i\bq\cdot\br},
\end{eqnarray}
and the inter-layer Coulomb interaction for the two layers separated by distance $\xi$ is 
\begin{eqnarray}
V^{\perp}(\br)&=&\frac{e^2}{\sqrt{\br^2+\xi^2}}=\int \frac{d^2\bq}{(2\pi)^2} \frac{2\pi e^2}{|\bq|}e^{-|\bq|\xi}e^{i\bq\cdot\br}.
\end{eqnarray}

To find the effective interaction within the twisted bilayer, we write the coherent state Feynman path integral action corresponding to $H_{int}$, Hubbard-Stratonovic decouple the intra-layer interaction in the Bernal bilayer, and integrate out the fermions in the Bernal bilayer and expand the result to quadratic order. Finally, we integrate out the Hubbard-Stratonovic field and obtain the static interaction
\begin{eqnarray}\label{Eqn:VeffSI}
V^{eff}(\br)&=&\int \frac{d^2\bq}{(2\pi)^2} \frac{2\pi e^2}{|\bq|}\left(1-e^{-2|\bq|\xi}\left(1-\frac{1}{\eps_{AB}(\bq)}\right)\right)e^{i\bq\cdot\br},
\end{eqnarray}
where the $\bq$-dependent dielectric constant of the Bernal bilayer can be related to its static polarization function $\Pi^0_{\bq}$ as $\eps_{AB}({\bq})=1+\frac{2\pi e^2}{|\bq|}\Pi^0_{\bq}$.

To gain a qualitative understanding of the above result, consider first the case when the Bernal bilayer is gated to a finite carrier density and thus acts as a metal. Then, $\Pi^0_{{\bq}\rightarrow 0}\rightarrow const.$ and the dielectric constant diverges as ${\bq}\rightarrow 0$. The effective interaction potential in momentum space then precisely corresponds to the real space potential produced by the test charge and its mirror image a distance $2\xi$ above the twisted bilayer as it should. On the other hand, when the Bernal bilayer is insulating, $\Pi^0_{{\bq}\rightarrow 0}\sim {\bq}^2$ and the long distance effective interaction is unchanged by the presence of the Bernal bilayer.

A more quantitative determination of the $V^{eff}(\br)$ requires a microscopic calculation of $\Pi^0_{\bq}$. We start with the low energy Hamiltonian for the electrons near the two valleys\cite{McCannPRB2006} labeled by $\tau_z=\pm 1$ acting on the basis $(\psi_{B2},\psi_{A1},\psi_{B1},\psi_{A2})$ as 
\begin{eqnarray}\label{Eqn:ABHamSI}
\mathcal{H}_{AB}({\bp})=\tau_z\left(\begin{array}{cccc}
-\frac{1}{2}\Delta & 0 & 0 &v_F(p_x-ip_y)\\
0 & \frac{1}{2}\Delta & v_F(p_x+ip_y) & 0 \\
0 & v_F(p_x-ip_y) & \frac{1}{2}\Delta & \tau_z \gamma_1 \\
v_F(p_x+ip_y) & 0 & \tau_z\gamma_1 & -\frac{1}{2}\Delta
\end{array}\right).
\end{eqnarray}
For polarization function we need
\begin{eqnarray}
\Pi^0_{\bq}(\Omega)=-k_BT\sum_{\omega_n}\int \frac{d^2\bk}{(2\pi)^2}\mbox{Tr}\left(G_{\bk+\bq}(\omega_n+\Omega_n)G_{\bk}(\omega_n)\right),
\end{eqnarray}
where the Matsubara frequency $\omega_n=(2n+1)\pi k_BT$ and the matrix Green's function is
\begin{eqnarray}
G_{\bk}(\omega_n)=\left(-i\omega_n 1_4+\mathcal{H}_{AB}(\bk)\right)^{-1}.
\end{eqnarray}
Because the $\mathcal{H}^2_{AB}$ is more easily invertable, we can find the $G_{\bk}(\omega_n)$ by first inverting $\omega^2_n+\mathcal{H}^2_{AB}$ and then left multiplying the result by $i\omega_n+\mathcal{H}_{AB}$. We find 
\begin{eqnarray}
G_{\bp}(\omega)&=&\sum_{s=\pm}\frac{i\omega 1_4+\mathcal{H}_{AB}(\bp)}{\omega^2+E^2_s(\bp)}\Lambda_s(\bp),
\end{eqnarray}
where the four eigenenergies are 
\begin{eqnarray}
E_s(\bp)&=&\pm\sqrt{\left(\frac{\Delta^2}{4}+\frac{\gamma^2_1}{2}+v^2_F\bp^2\right)+s\sqrt{\frac{\gamma^4_1}{4}+(\gamma^2_1+\Delta^2)v^2_F{\bp}^2}},
\end{eqnarray}
for $s=\pm 1$, and 
\begin{eqnarray}
\Lambda_s(\bp)&=&\frac{1}{2}\left(1_4-\frac{s}{\sqrt{\frac{\gamma^4_1}{4}+(\gamma^2_1+\Delta^2)v^2_F{\bp}^2}}\left(\begin{array}{cc}
\frac{\gamma^2_1}{2}  & -v_F{\bp}\cdot\sigma \left(\gamma_1 \tau_z\sigma_1+\Delta\sigma_3\right)\\
-\left(\gamma_1 \tau_z\sigma_1+\Delta\sigma_3\right)v_F{\bp}\cdot\sigma &  -\frac{\gamma^2_1}{2}
\end{array}\right)\right).
\end{eqnarray}
Finally, including the spin and valley degeneracy:
\begin{eqnarray}\label{Eqn:PiSI}
\Pi^0_\bq=\sum_{\tau_z=\pm}\sum_{s_1,s'_1=\pm}\sum_{s_2,s'_2=\pm}\int\frac{d^2\bp}{(2\pi\hbar)^2}\frac{\tanh\left(\frac{s_1'E_{s_1}(\bp+\bq)-\mu}{2k_BT}\right)-
\tanh\left(\frac{s_2'E_{s_2}(\bp)-\mu}{2k_BT}\right)
}{s_1'E_{s_1}(\bp+\bq)-s_2'E_{s_2}(\bp)}
\mbox{Tr}\left[P_{s_1,s_1'}(\bp+\bq) P_{s_2,s_2'}(\bp)\right],
\end{eqnarray}
where the projector on the individual eigenstates is
\begin{eqnarray}
P_{s,s'}(\bp)&=&\frac{1}{2}\left(1+s'\frac{\mathcal{H}_{AB}(\bp)}{E_{s}(\bp)}\right)\Lambda_s(\bp).
\end{eqnarray}
We can rescale all the energies by $\gamma_1$ which allows us to rewrite the polarization function via a scaling function $\Phi$ as 
\begin{eqnarray}
\Pi^0_{\bq}&=&\frac{\gamma_1}{\hbar^2v^2_F}\Phi\left(\frac{\hbar v_F q}{\gamma_1},\frac{\Delta}{\gamma_1},\frac{\mu}{\gamma_1},\frac{k_BT}{\gamma_1}\right).
\end{eqnarray}
We determine $\Phi$ by numerically performing the integral (\ref{Eqn:PiSI}).
The static dielectric function is then
\begin{eqnarray}
V_\bq\Pi^0_{\bq}(0)&=&2\pi \frac{e^2}{\eps_{hBN}|\bq|}\frac{\gamma_1}{\hbar^2v^2_F}\Phi\left(\frac{\hbar v_F q}{\gamma_1},\frac{\Delta}{\gamma_1},\frac{\mu}{\gamma_1},\frac{k_BT}{\gamma_1}\right)
=\frac{2\pi}{\eps_{hBN}}\times\frac{c}{v_F}\times \frac{e^2}{\hbar c}\times\frac{\gamma_1}{\hbar v_F q}\Phi\left(\frac{\hbar v_F q}{\gamma_1},\frac{\Delta}{\gamma_1},\frac{\mu}{\gamma_1},\frac{k_BT}{\gamma_1}\right).
\end{eqnarray}
Writing the exponential screening factor in Eqn.(\ref{Eqn:VeffSI})  using scaling variables, we get
$e^{-2|\bq|\xi}=\exp\left(-\left(\frac{\hbar v_F q}{\gamma_1}\right)\left(2\frac{\gamma_1\xi}{\hbar v_F}\right)\right)$.
For $\xi=3$nm, $\gamma_1=0.3eV$, and $v_F=10^6m/s$, we have $2\frac{\gamma_1\xi}{\hbar v_F}=2.735$; we also set $\eps_{hBN}=4.4$.
In the Figure \ref{Fig:ScreeningCNP} we place the chemical potential $\mu$ at the CNP and vary the gap $\Delta$ of the Bernal bilayer from $0$meV to $30$meV. We see that the reduction of the repulsive interaction in the twisted bilayer is minimized when the CNP gap of the Bernal bilayer is maximized.
In Figure \ref{Fig:ScreeningFilling} we set $\Delta=30$meV and vary the carrier concentration, moving the chemical potential from the middle of the gap to $15$meV above the conduction band minimum. We see that the reduction of the repulsive interaction in the twisted bilayer is minimized when the chemical potential sits in the gap of the Bernal bilayer. We also see that the interaction is almost independent of the filling of the Bernal bilayer once it is conducting. Remarkably, the maximal increase in the superconducting $T_c$ as well as the gap of the correlated insulator correspond to the strongest Coulomb repulsion $V^{eff}_\bq$ i.e. the minimal reduction of $V^{\parallel}_\bq$. 
\begin{figure}[h]
	\centering
	
		\includegraphics[width=1\linewidth]{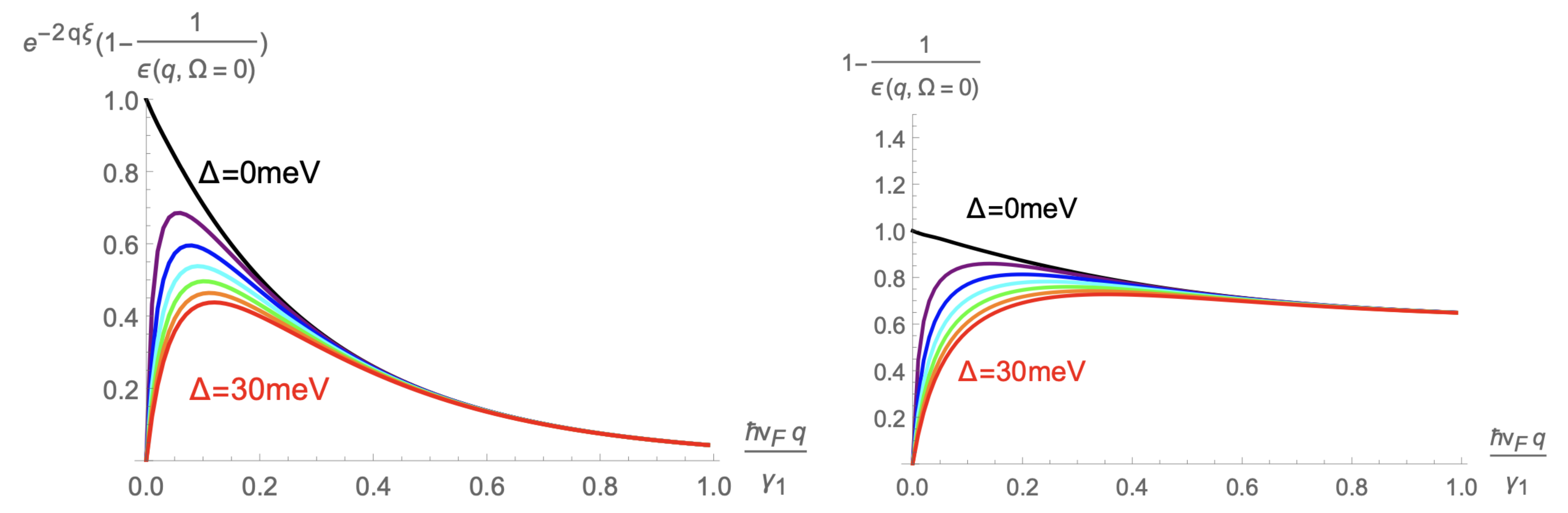}
	\caption{The effective interaction among the electrons in the twisted bilayer due to static screening within the AB stacked Bernal bilayer is $
V^{eff}_\bq
=V_\bq^\parallel
\left(1-e^{-2|\bq|\xi}\left(1-\frac{1}{\eps_{AB}(\bq)}\right)\right)
 $.
Left panel: the reduction factor as a function of dimensionless wavevector $q$ at low temperature; $\gamma_1\approx 0.3$eV is the direct interlayer tunneling defined in Eqn.(\ref{Eqn:ABHamSI}).
 The series of curves shows the evolution from semimetallic Bernal bilayer $\Delta=0$ (top curve) to progressively more insulating Bernal bilayer as $\Delta$ is varied in steps of $5$meV, ending with the Bernal gap at $30$meV (bottom curve).
Right panel: The dielectric reduction factor without the exponential.
}
	\label{Fig:ScreeningCNP}
\end{figure}

\begin{figure}[h]
	\centering
	\includegraphics[width=1\linewidth]{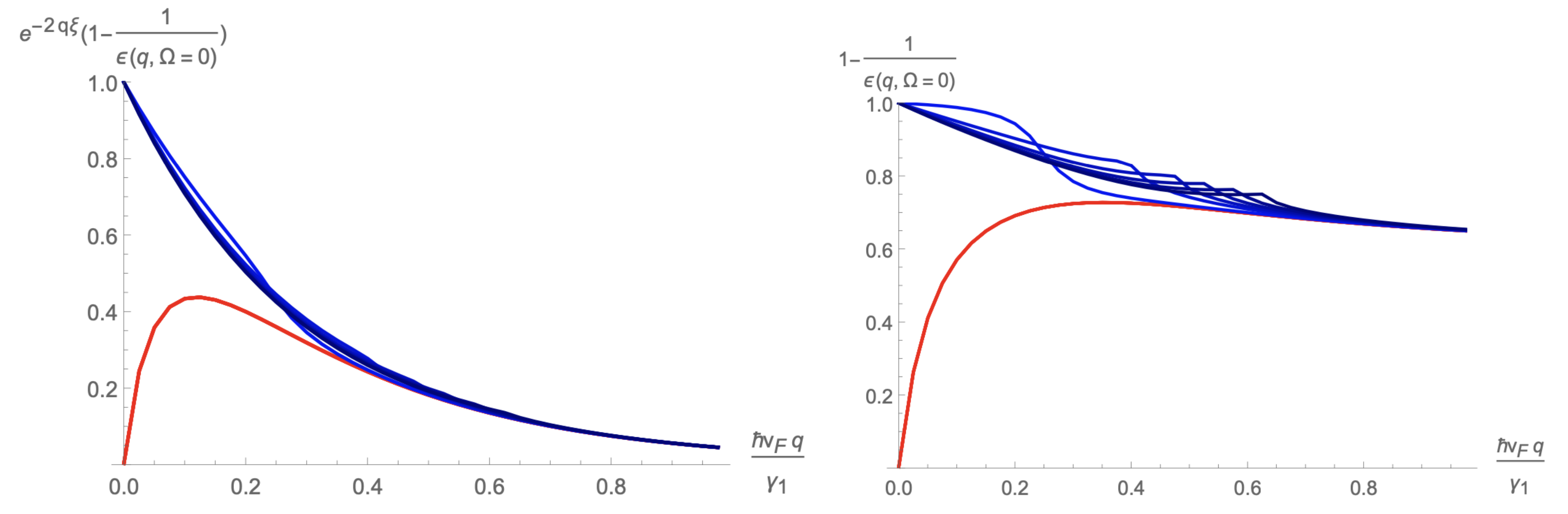}
	\caption{The effective interaction among the electrons in the twisted bilayer due to static screening within the AB stacked Bernal bilayer is $
V^{eff}_\bq
=V_\bq^\parallel
\left(1-e^{-2|\bq|\xi}\left(1-\frac{1}{\eps_{AB}(\bq)}\right)\right)
 $.
Left panel: the reduction factor as a function of dimensionless wavevector $q$ at low temperature compared to the gap.
 The series of curves shows the evolution from the insulating Bernal bilayer (red curve) to conducting (blue), as the chemical potential is varied in steps of $3$meV, starting from being in the middle of the Bernal gap (set to $30$meV), to being $15$meV above.
Right panel: the dielectric reduction factor without the exponential; the $2k_F$ features in the metallic curves are also visible.
}
	\label{Fig:ScreeningFilling}
\end{figure}

\subsection{Sample fabrication}

\begin{figure}
\includegraphics[width=0.65\linewidth]{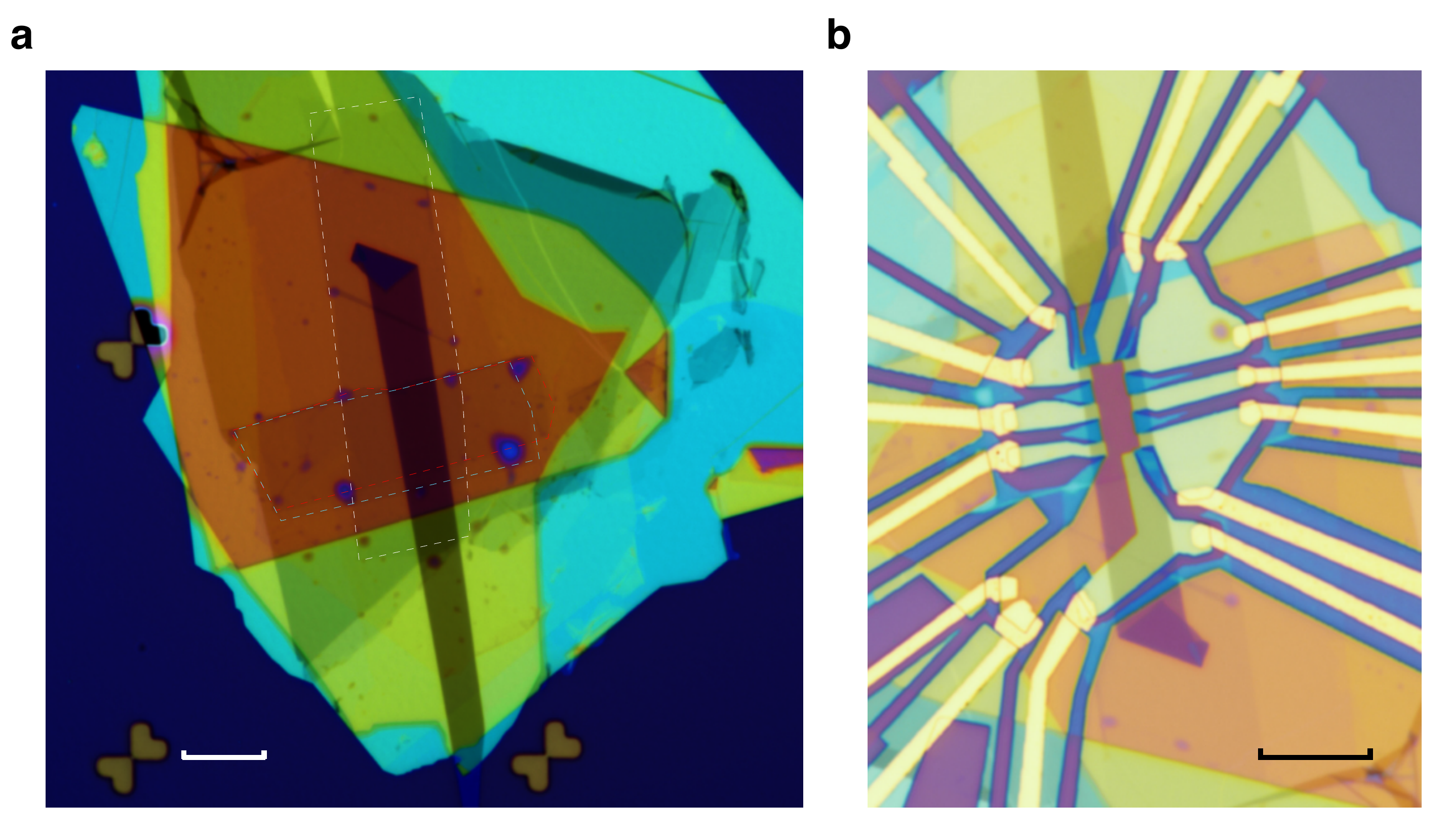}
\caption{\label{fig:stack} { Optical image of the hybrid double-layer structure (a) before and (b) after nanofabrication. The Bernal bilayer graphene layer is highlighted by the white dashed contour, whereas two graphene layers in tBLG are highlighted by the red and blue contours, respectively. The scale bars correspond to $10 \mu$m. }}
\end{figure}

\begin{figure}
\includegraphics[width=0.4\linewidth]{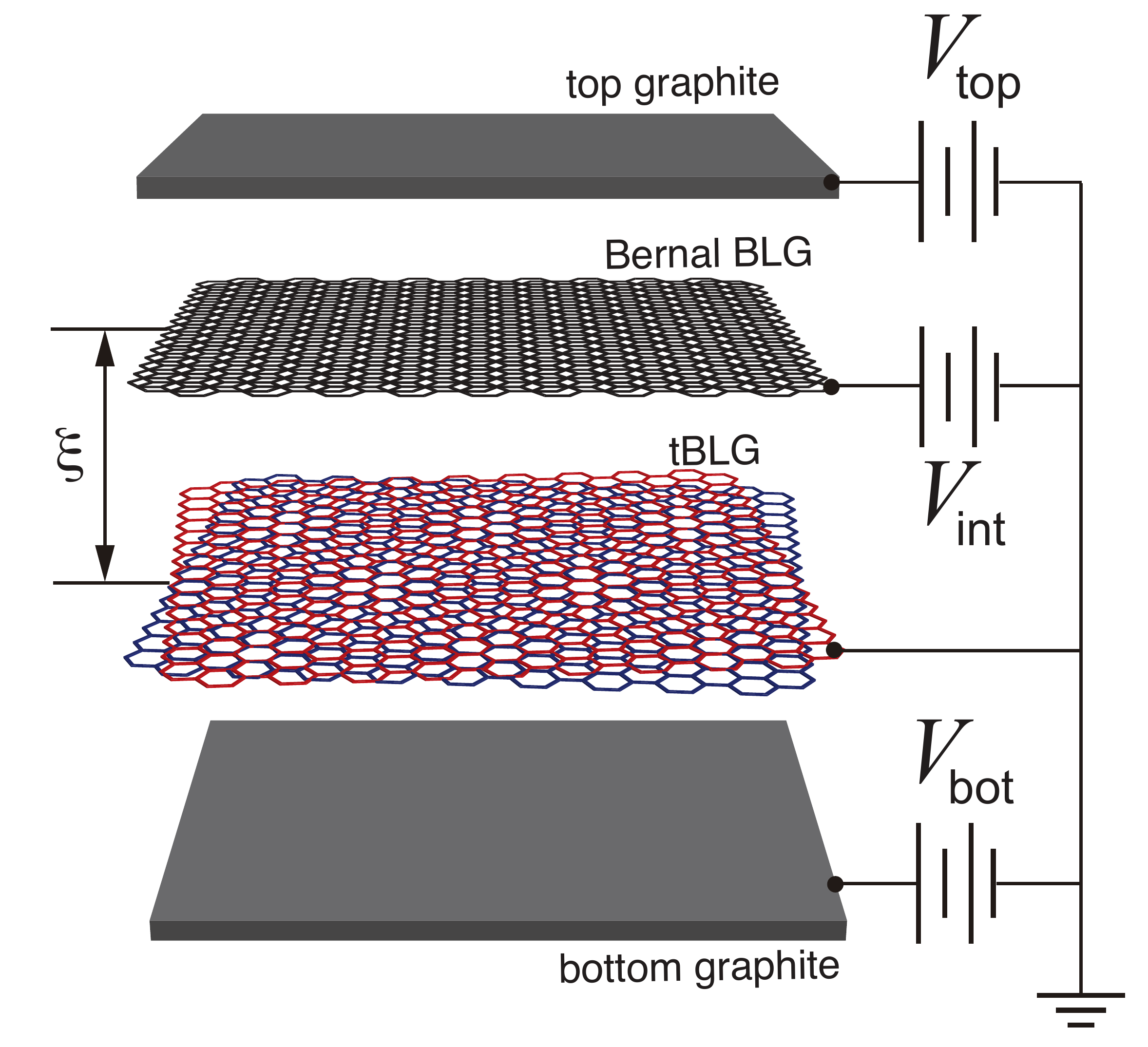}
\caption{\label{fig:gate} {\bf{Schematic diagram for transport measurement}} Transport measurement is performed with DC voltage bias applied to top graphite, bottom graphite and Bernal BLG, to control charge carrier density in BLG and tBLG, as well as displacement field $D$. }
\end{figure}

The tBLG in the hybrid double-layer structure is assembled using the ``cut-and-stack'' technique ~\cite{Saito2019decoupling}, instead of the ``tear-and-stack'' technique ~\cite{Kim2016twist,Cao2018a}. All components of the structure are assembled from top to bottom using the same poly(bisphenol A carbonate) (PC)/polydimethylsiloxane (PDMS) stamp mounted on a glass slide. The sequence of stacking is: graphite as top gate electrode, $30$ nm thick hBN as top dielectric, Bernal bilayer graphene, $3$ nm thick hBN as insulating barrier, magic-angle tBLG, $30$ nm thick hBN as bottom dielectric, bottom graphite as bottom gate electrode. The entire structure is deposited onto a doped Si/SiO$_2$ substrate, as shown in Fig.SI~\ref{fig:stack}a. Electrical contact to both Bernal and twisted bilayers are made independently by CHF$_3$/O$_2$ etching and deposition of the $Cr/Au$ (2/100 nm) metal edge contact.

The hybrid double-layer sample is shaped into an aligned Hall bar geometry. In this geometry, each layer has independent electrical contact for longitudinal and Hall voltage measurements, as shown in Fig.SI~\ref{fig:stack}b.

\subsection{Transport measurement}

\begin{figure}
\includegraphics[width=0.85\linewidth]{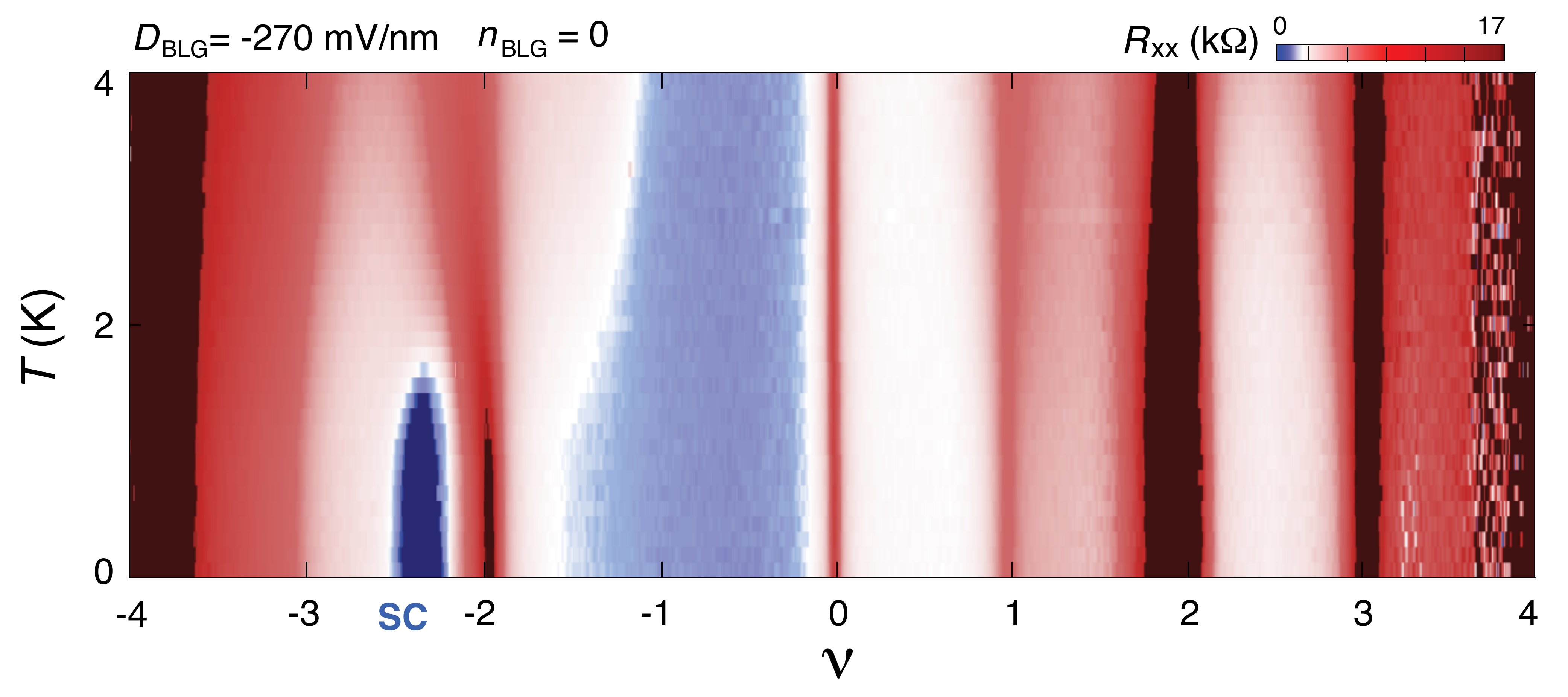}
\caption{\label{fig:SCT} {$R_{xx}$ versus $\nu_{tBLG}$ and $T$ measured from the tBLG at $n_{BLG} = 0$ and $D_{BLG}=-270$ mV/nm. We note that only one superconducting pocket is observed in our sample on the hole doping side of $\nu=-2$, which is similar to previous observations in reference ~\cite{Saito2019decoupling} where Coulomb screening from nearby graphite layer is weak. Based on this, the absence of other superconducting pockets is likely associated with sample details, such as twist angle inhomogeneity, and we do not think it results from Coulomb screening from a nearby BLG. }}
\end{figure}

%\begin{figure}
%\includegraphics[width=0.88\linewidth]{FigSI2.pdf}
%\caption{\label{fig:RT} {(a) $R_{xx}$ versus $T$ for the superconducting phase at optimal doping, measured at $n_{BLG} = 0$ with different $D$. (b) $R_{xx}$ as a function of $T$ and $n_{tBLG}$, measured at $n_{BLG} = 0$ with different $D$. }}
%\end{figure}

The device geometry of the hybrid double-layer structure allows independent control of carrier density in Bernal BLG and tBLG, $n_{BLG}$ and $n_{tBLG}$, as well as displacement field $D$. Such control is achieved by applying a DC gate voltage to top graphite electrode $V_{top}$, bottom graphite electrode $V_{bot}$, along with a voltage bias between BLG and tBLG $V_{int}$. $n_{BLG}$, $n_{tBLG}$ and $D$ can be obtained using the following equations: 
\begin{eqnarray}
n_{BLG} &=& (C_{top}V_{top}+C_{int}V_{int})/e+n^0_{BLG}, \label{EqM1}\\
D_{BLG} &=& (C_{top}V_{top}-C_{int}V_{int})/\epsilon_0, \label{EqM2}\\
n_{tBLG} &=& (C_{bot}V_{bot}+C_{int}V_{int})/e+n^0_{tBLG}, \label{EqM3}\\
D_{tBLG} &=& (-C_{bot}V_{bot}+C_{int}V_{int})/\epsilon_0, \label{EqM4}
\end{eqnarray} 
\noindent
where $C_{top}$ is the geometric capacitance between top graphite and BLG, $C_{bot}$ the geometric capacitance between bottom graphite and tBLG, and $C_{int}$ the geometric capacitance between BLG and tBLG. $n^0_{BLG}$ and $n^0_{tBLG}$ are intrinsic doping in BLG and tBLG, respectively. 

Transport measurement is performed in a BlueFors LD400 dilution refrigerator with a base temperature of $15$ mK. Temperature is measured using a resistance thermometer located on the sample probe. Standard low frequency lock-in techniques with Stanford Research SR830 amplifier are used to measure resistance $R_{xx}$ and $R_{xy}$, with an excitation current of $0.6-1$ nA at a frequency of $17.77-43.33$ Hz. An external multi-stage low-pass filter is installed on the mixing chamber of the dilution unit. The filter contains two filter banks, one with RC circuits and one with LC circuits. The radio frequency low-pass filter bank (RF) attenuates above $80$ MHz, whereas the low frequency low-pass filter bank (RC) attenuates from $50$ kHz. The filter is commercially available from QDevil.

%$D$-field dependenceWe note that the $D$-field has little to no consequence on the transport behavior of tBLG based on the CIs, which is consistent with the theoretical prediction of strong interlayer hybridization ~\cite{Bistritzer2011,Trambly2016,Kim2017tunable}. This insensitivity eliminates $D$-dependence in transport measurement arising from disorder concentrated on one graphene layer ~\cite{Yankowitz2019SC}. 
Both BLG and tBLG in the hybrid double-layer structure exhibit charge carrier inhomogeneity of $\delta n < \pm 2 \times 10^{10}$, measured by the divergent Hall resistance peaks at a small magnetic field $B = 0.15$ T. The excellent sample quality offers a pristine environment to examine the effect of Coulomb screening with high sensitivity.

%\begin{figure}
%\includegraphics[width=0.55\linewidth]{FigHNfit.pdf}
%\caption{\label{fig:HN} {R  }}
%\end{figure}

\begin{figure}
\includegraphics[width=0.75\linewidth]{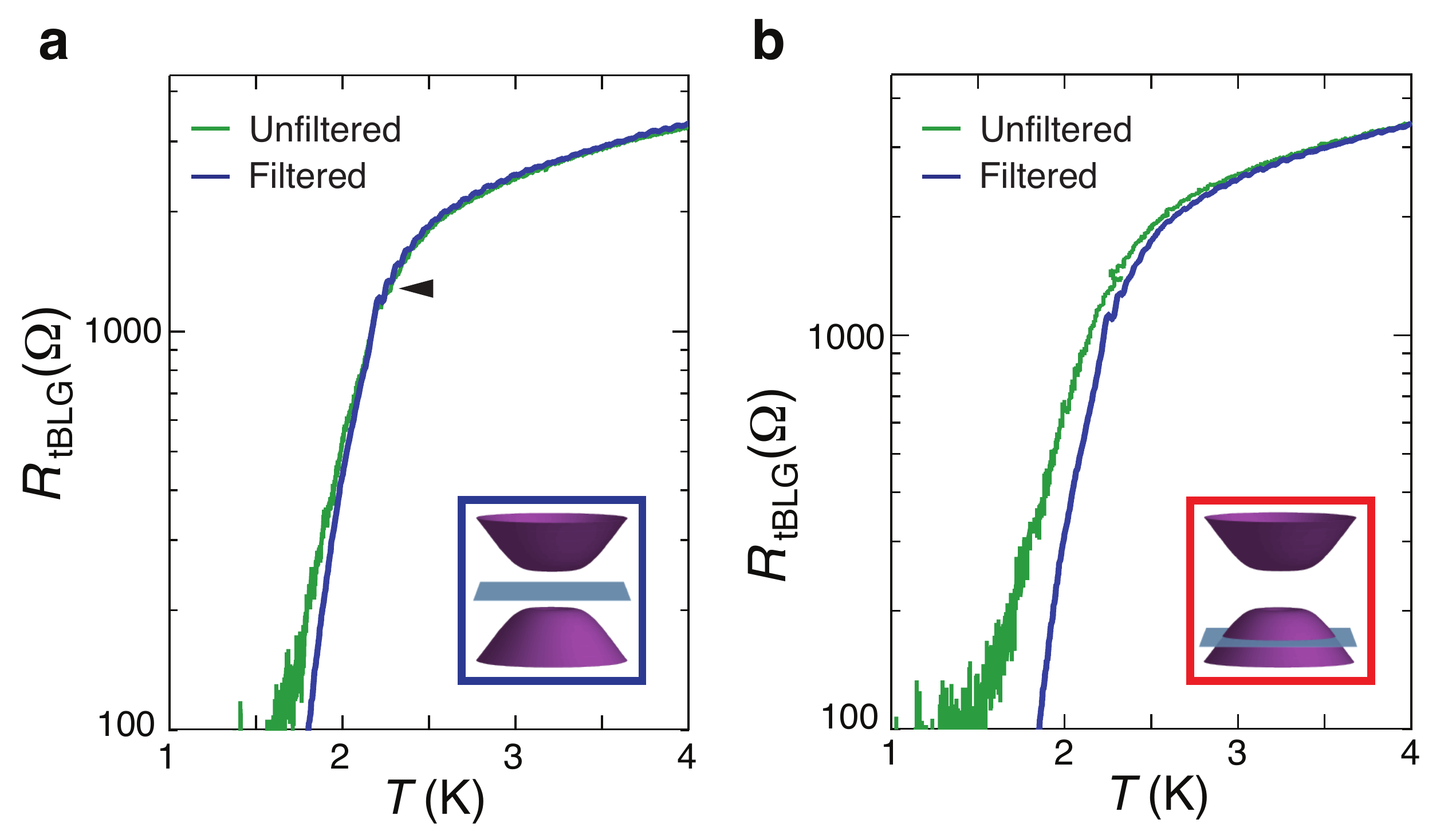}
\caption{\label{fig:SCT} {The influence of low-pass filter. $R_{tBLG}$ as a function of temperature, showing the superconducting-normal state transition measured at different $n_{BLG}$ and $D=-350$ mV/nm. Blue traces are measured with, and green trace without filtering. The measurements are done when BLG is (a) fully insulating at $n_{BLG}=0$ and (b) metallic with $n_{BLG}=-0.5 \times 10^(12)$ cm$^{-2}$.  (a) When BLG is fully insulating, the normal state resistance is not influenced by filtering. The unfiltered trace deviates slightly from the filteredd trace at temperature below $T_c$. (b) When BLG is metallic, the unfiltered trace starts to deviate from the filteredd trace at temperature above $T_c$, resulting in a significantly suppressed $T_c$ value.  }}
\end{figure}

%\begin{figure}
%\includegraphics[width=0.58\linewidth]{FigSI8v3.pdf}
%\caption{\label{fig:CI} { $R_{xx}$ as a function of $\n_{tBLG}$ and $n_{BLG}$, measured at $D_{BLG} = -90$ mV/nm and $T = 15$ mK.  Right inset: schematic for energy band of BLG, with different charge carrier density. BLG is insulating in the density range between two white dashed line. In this regime, screening from BLG is absent, tBLG is tuned with both top and bottom graphite gate, which gives rise to the distortion in both insulating and superconducting features. Most importantly, both the CI and the superconducting phases are more robust when  is highlighted energy gap of BLG is  }}
%\end{figure}

%\begin{figure}
%\includegraphics[width=0.7\linewidth]{FigSI10RT.pdf}
%\caption{\label{fig:RT2} {$R_{xx}$ versus $T$ for the superconducting phase at optimal doping, measured at (a) $D_{BLG}=-50$ mV/nm and different $n_{BLG}$; (b) and (c) $n_{BLG} = 0$ and different $D_{BLG}$. $R_{xx}$ is normalized to its value at $T = 5$ K in panel (b). The systematic shift in the temperature dependence of $R_{xx}$ indicates that the relative change in $T_c$ is independent of How $T_c$ is defined. }}
%\end{figure}

\begin{figure}
\includegraphics[width=1\linewidth]{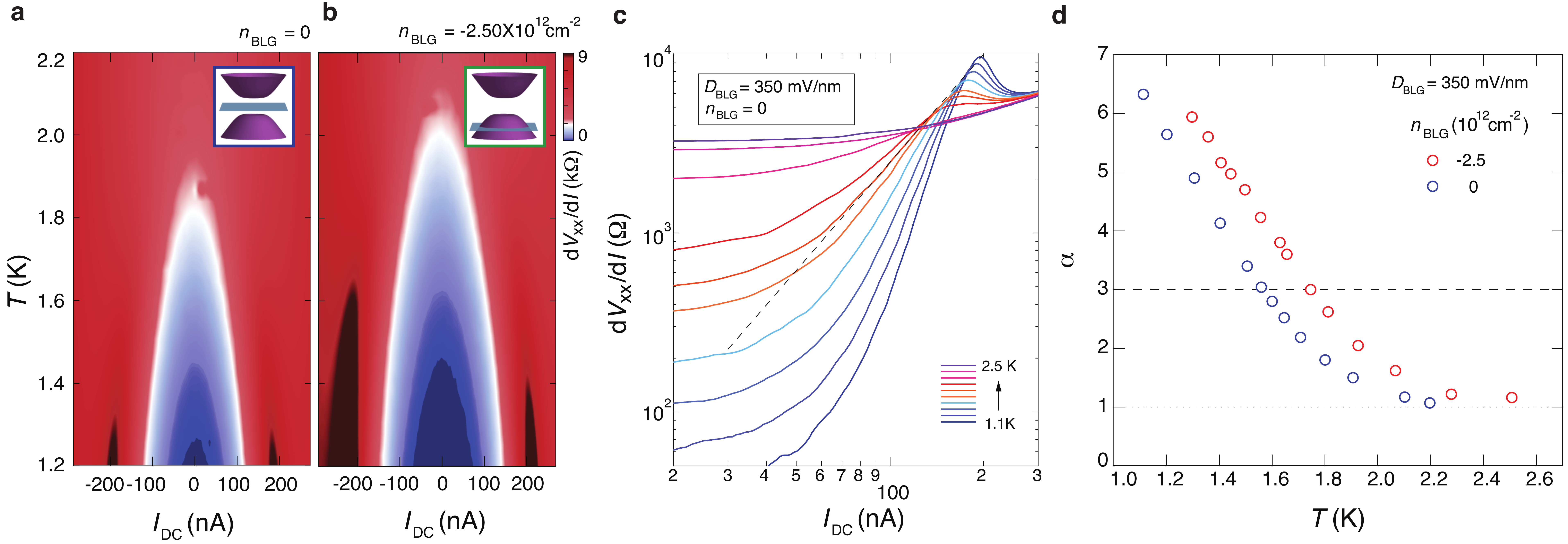}
\caption{\label{fig:IV} {Differential resistance $dV_{xx}/dI$ versus d.c. bias current $I$ and $T$, measured at (a) $n_{BLG}=0$ and (b) $-2.5\times 10^{12}$ cm$^{-2}$. A large energy gap in BLG is induced by $D = 350$ mV/nm. tBLG is tuned to optimal doping  $n_{tBLG} = -1.5 \times 10^{12}$ cm$^{-2}$. (c) $dV_{xx}/dI$ versus d.c. bias in log scale measured at $D = 350$ mV/nm. The black dashed line is a fit to the $V = I^3$ power law, indicating a BKT type transition. (d) Power coefficient $\alpha$ extracted from panel (c) as a function of T. A BKT type transition is defined by $\alpha = 3$. Tuning density carrier in BLG has a measurable influence on the BKT transition in tBLG. }}
\end{figure}

\begin{figure}
\includegraphics[width=0.8\linewidth]{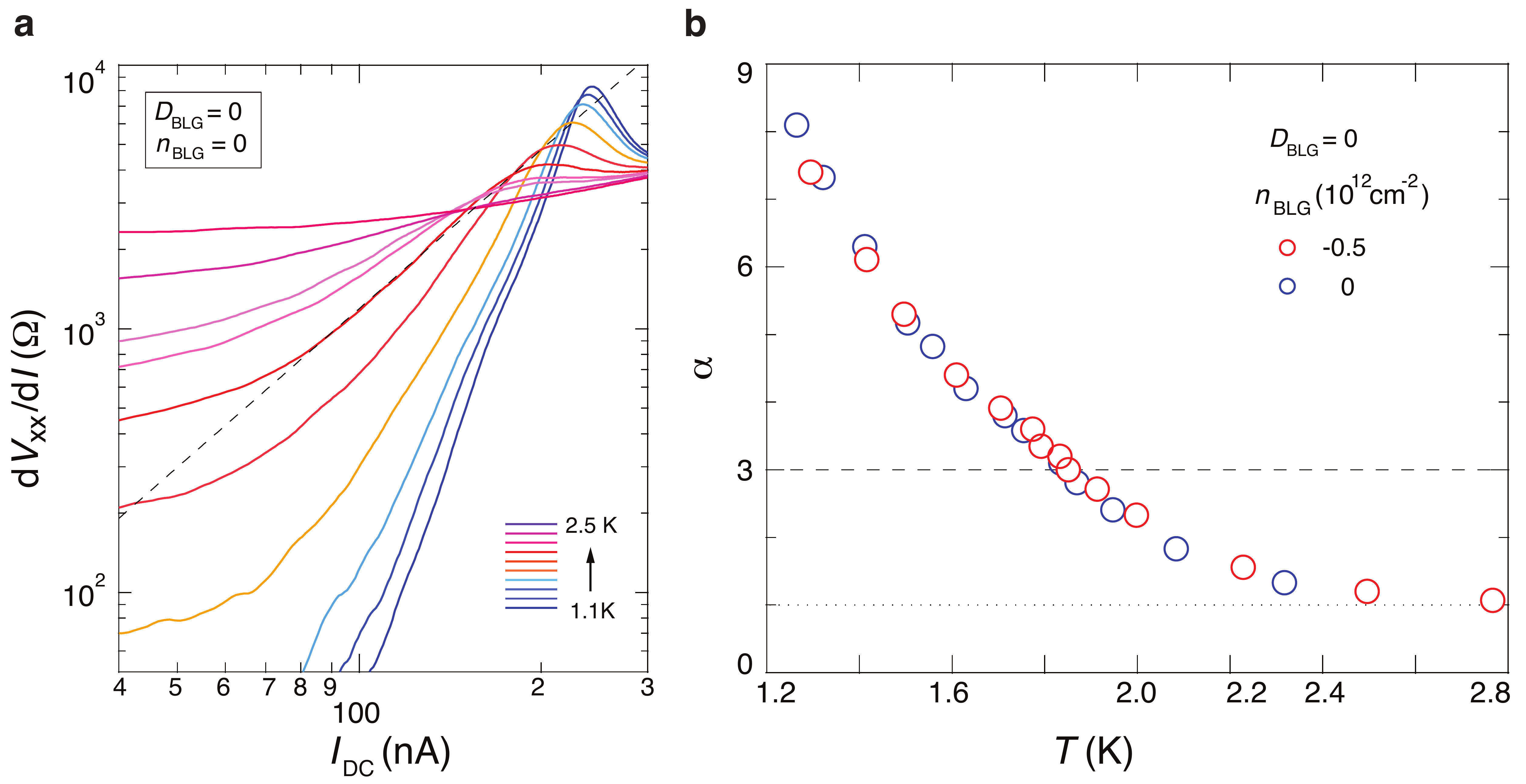}
\caption{\label{fig:IV0mV} {(a) $dV_{xx}/dI$ versus d.c. bias in log scale measured at $D = 0$. The black dashed line is a fit to the $V = I^3$ power law, indicating a BKT type transition. (b) Power coefficient $\alpha$ extracted from panel (a) as a function of $T$. At $D = 0$, the effect of tuning density carrier in BLG on the superconducting phase in tBLG is negligible. }}
\end{figure}

\begin{figure}
\includegraphics[width=0.5\linewidth]{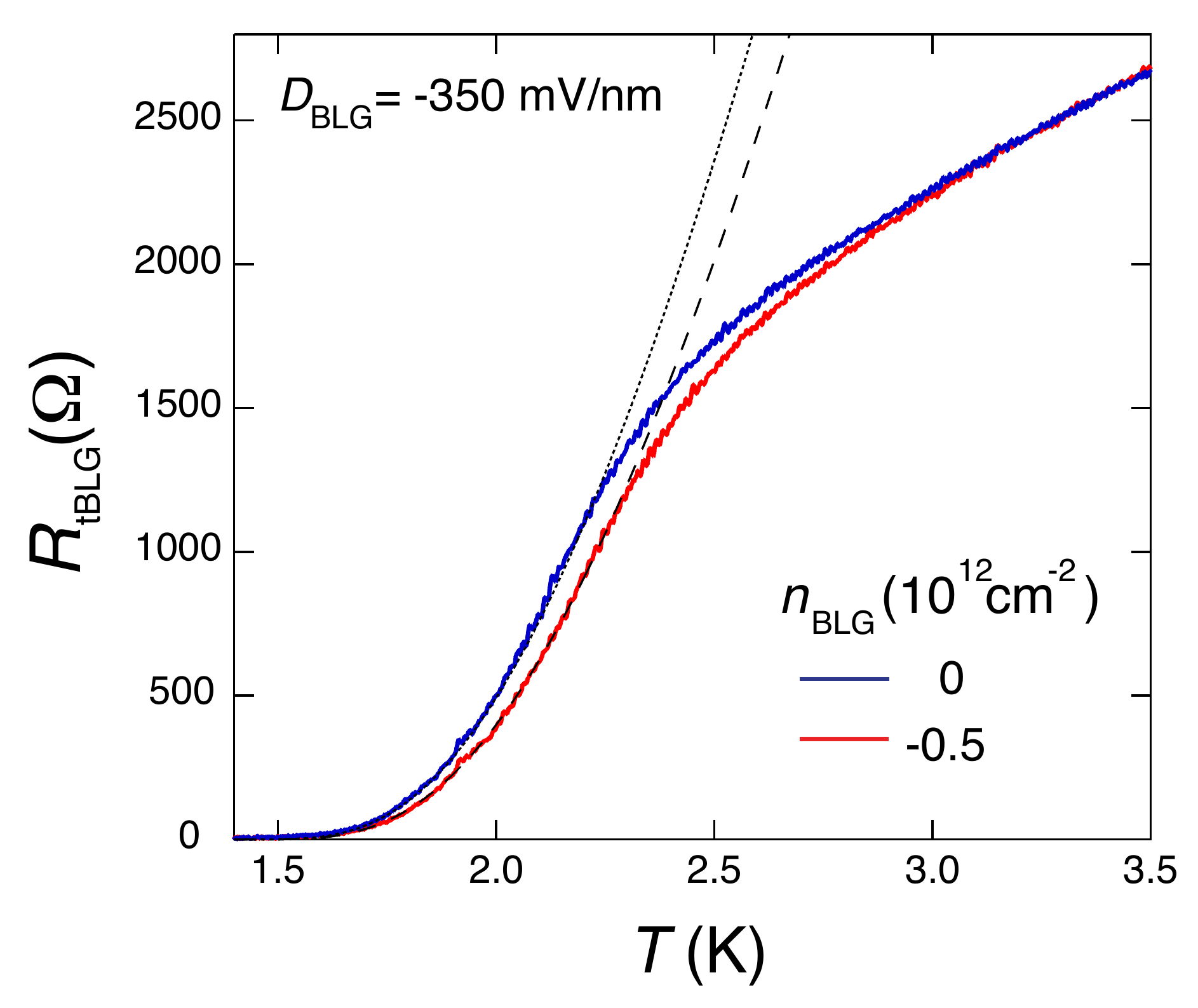}
\caption{\label{fig:HN} {(a) $R_{tBLG}$ as a function of $T$ measured at $D = -350$ mV/nm with different $n_{BLG}$. The BKT temperature can be obtained using Halperin-Nelson form $R(T) = R_0 exp[-b/(T -T_{BKT})^{1/2}]$ (black dashed and dotted trace). Using the H-N fit, $T_{BKT}$ is $1.38$ K for $n_{BLG}=0$ and $1.46$ K for $n_{BLG} = -0.5\times10^{12}$ cm$^{-2}$.
}}
\end{figure}

\begin{figure}
\includegraphics[width=0.95\linewidth]{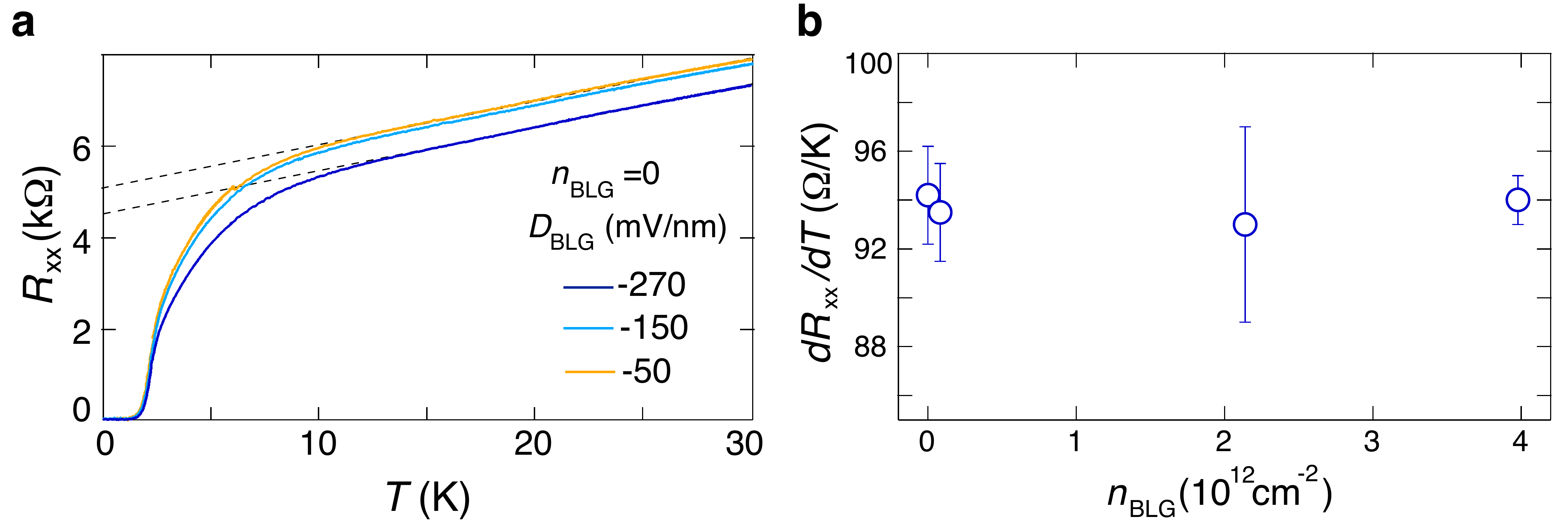}
\caption{\label{fig:slope} {(a) $R_{tBLG}$ as a function of $T$ measured at optimal doping, showing a linear in $T$ behavior at $T > T_c$ at $n_{BLG} = 0$ with varying $D_{BLG}$. (b) The slope of $R_{tBLG}$ in the T-linear regime as a function of $n_{BLG}$ at $D_{BLG} = -150$ mV/nm. When BLG is tuned over this range of parameters, apparent variations are observed in the CI and superconducting states. In comparison, there is no detectable variation in $dR_{xx}/dT$. }}
\end{figure}

\begin{figure}
\includegraphics[width=0.65\linewidth]{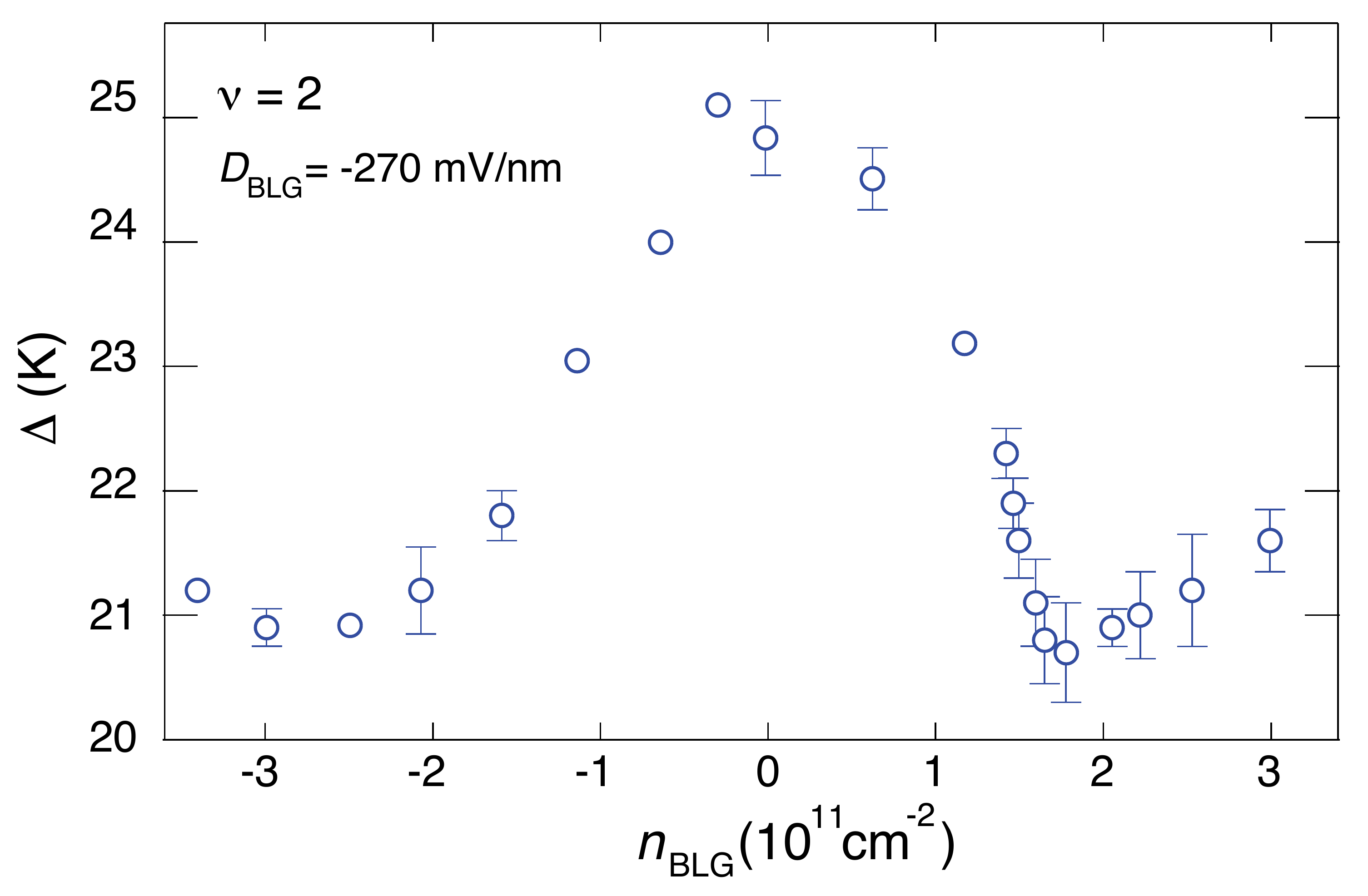}
\caption{\label{fig:CI} {
%(a) Arrhenius plot for the CI at $\nu=-2$ measured at $D_{BLG} = -90$ mV/nm with different carrier density in BLG. A larger energy gap is observed when BLG is insulating with $n_{BLG}=0$, compared to when BLG is conductive at $n_{BLG} = -2.3 \times 10^{11}$ cm$^{-2}$. 
$\Delta_{\nu=2}$ as a function of $n_{BLG}$ measured at $D_{BLG} = -270$ mV/nm. In the presence of a large energy gap induced by $D_{BLG}$, The energy gap for CIs at $\nu=2$ is significantly suppressed as BLG transitions from fully insulating to metallic. }}
\end{figure}

\begin{figure}
\includegraphics[width=0.5\linewidth]{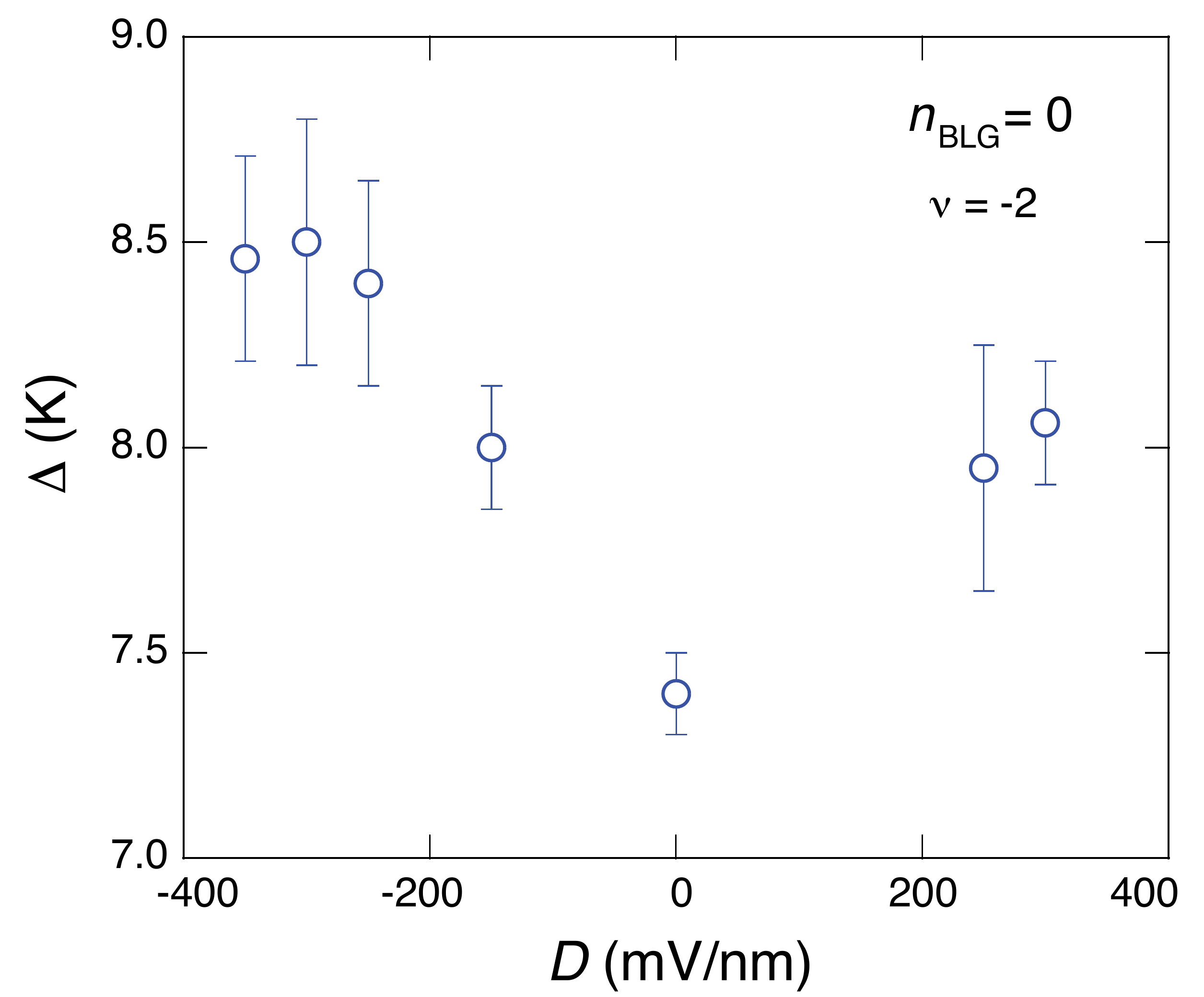}
\caption{\label{fig:m2} {$\Delta_{\nu=-2}$ as a function of $D_{BLG}$ measured at $n_{BLG}=0$. When BLG is insulating at large $D$, a larger energy gap is observed for the CI at $\nu=-2$, compared to the situation where BLG is a semi-metal at $D=0$.  }}
\end{figure}

\begin{figure}
\includegraphics[width=0.55\linewidth]{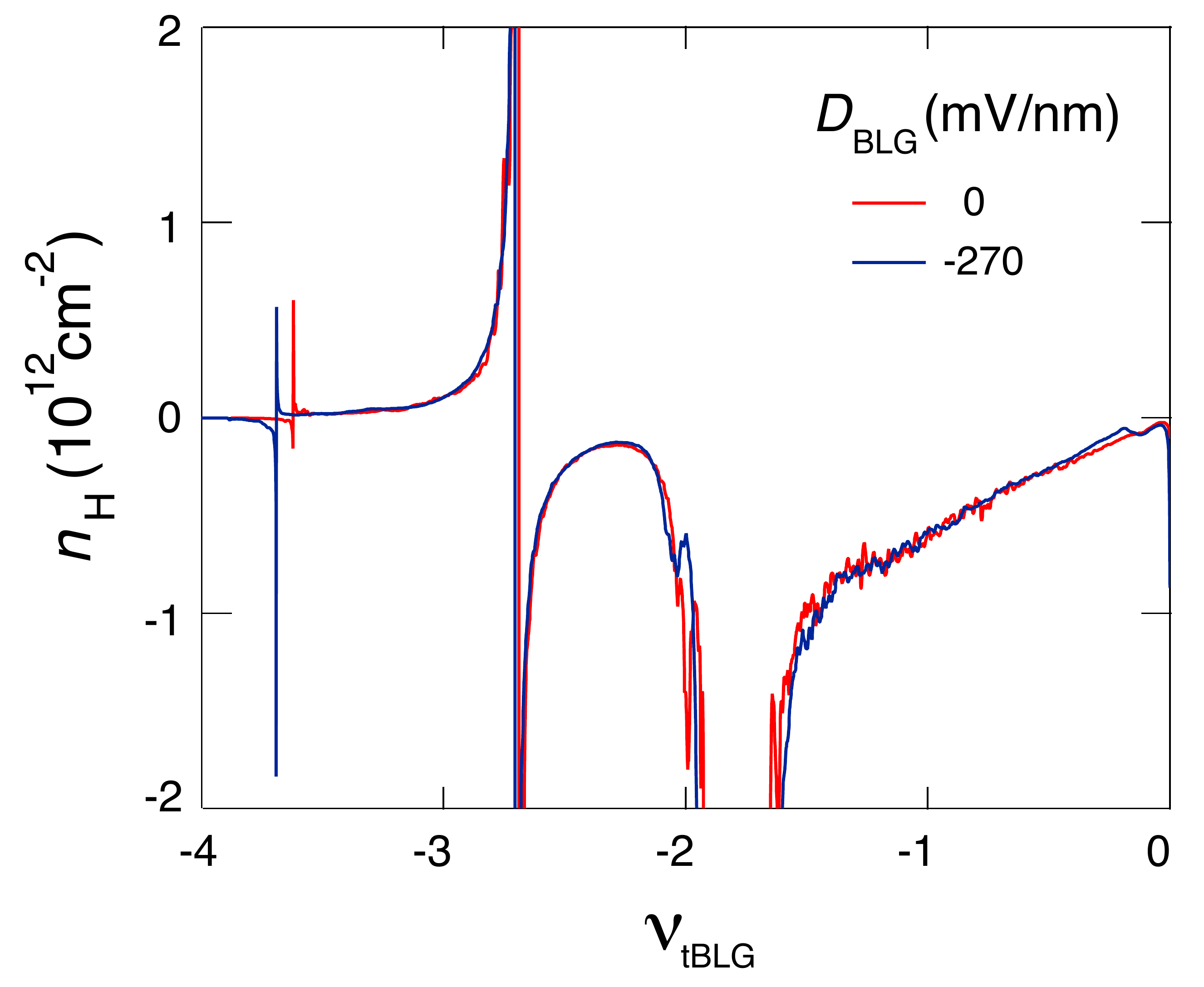}
\caption{\label{fig:Hall} {Hall density $n_H$ as a function of $\nu_{tBLG}$ measured at $B = 0.15$ T, $n_{BLG}=0$ with different $D_{BLG}$. The Hall density resets to zero at $\nu =-2$, indicating the formation of a new, small Fermi surface. The behavior of Hall density is insensitive to Coulomb screening from BLG, suggesting that the tunability to the CI and the superconducting phase does not result from changes in Fermi surface reconstruction near $\nu=-2$. }}
\end{figure}

\begin{figure}
\includegraphics[width=0.9\linewidth]{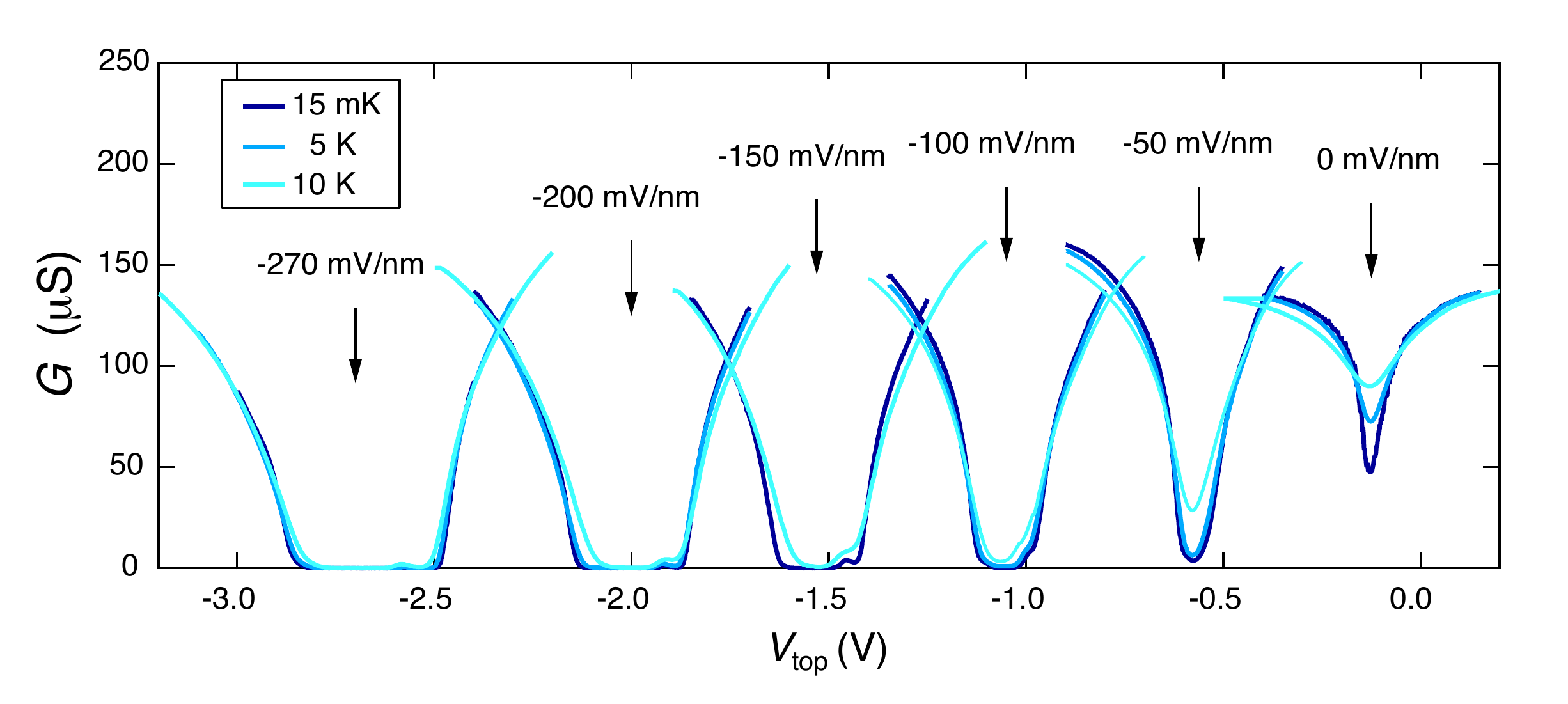}
\caption{\label{fig:BLG} {Conductance of BLG as a function of $V_{top}$ measured at different $D$ and $T$. For $|D| > 16$ mV/nm, the conductance of BLG remains constant up to $5$ K. For $|D| > 30$ mV/nm, the conductance of BLG remains constant up to $10$ K. As a result, $\Delta_{\nu=2}$ and $T_c$ measurements in tBLG reported in this work are not influenced by the temperature dependence of BLG. }}
\end{figure}

\begin{figure}
\includegraphics[width=0.5\linewidth]{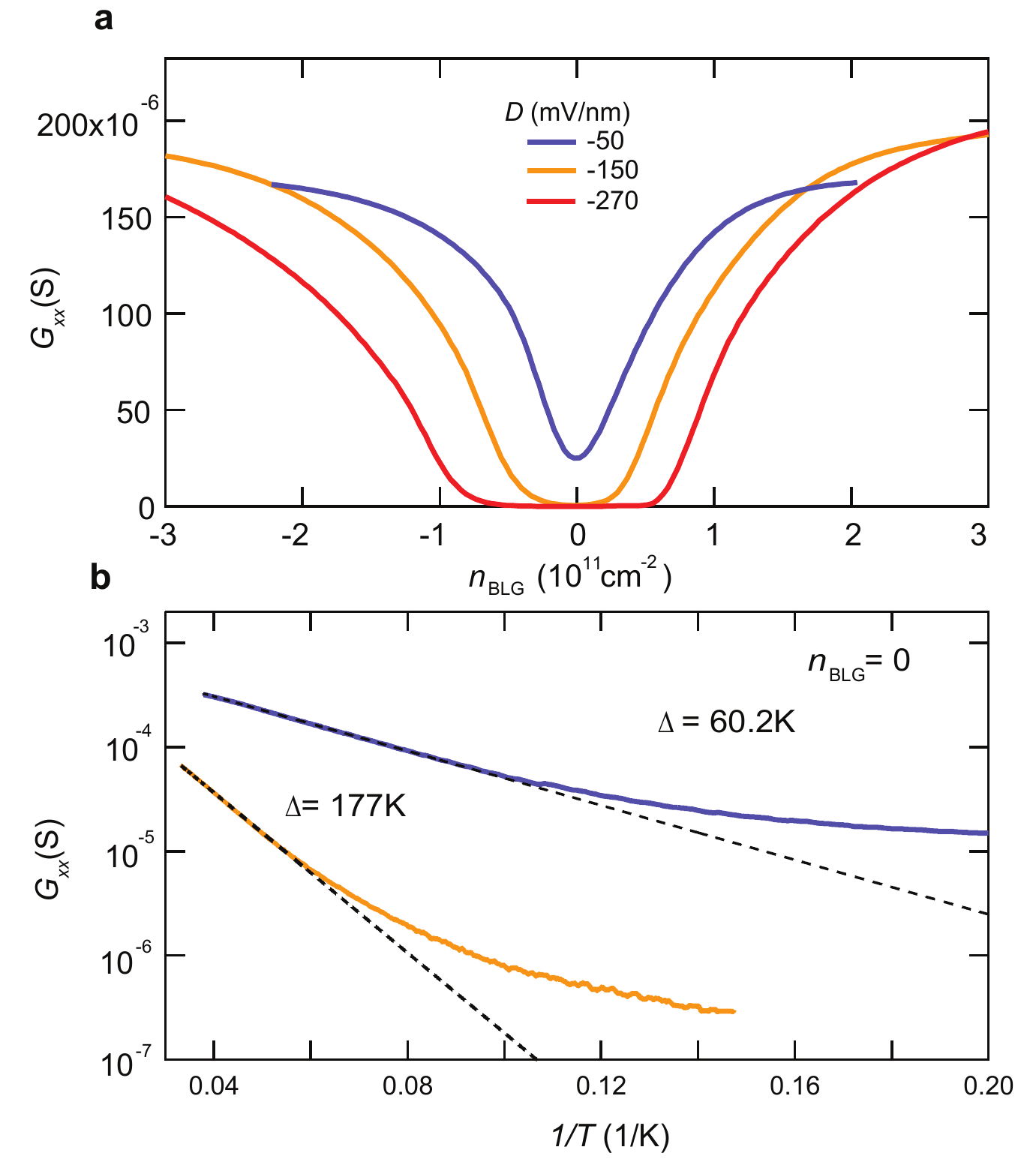}
\caption{\label{fig:BLG2} {(a) Conductance of BLG as a function of $n_{BLG}$ measured at different $D$ and $T = 15$ mK. (b) Arrhenius plot measured in BLG, showing the amplitude of energy gap near the CNP at different $D_{BLG}$. The energy gap at $D = 90$ mV/nm is much larger than $177$ K. It is not shown here due to the high temperature range required for such measurement. }}
\end{figure}

%\begin{figure}
%\includegraphics[width=0.6\linewidth]{FigSI4.pdf}
%\caption{\label{fig:SC} {Differential resistance $dV_{xx}/dI$ versus d.c. bias current $I$ measured at different $D_{BLG}$ with $n_{BLG}=0$.  }}
%\end{figure}

\newpage

\end{widetext}

\end{document}